\documentclass[12pt, a4paper]{article}

\usepackage{graphicx}
\usepackage{bmpsize}
\usepackage{amsmath}

\DeclareMathOperator{\arctanh}{arctanh}

\DeclareMathOperator{\arccoth}{arcCoth} 
\usepackage{amsfonts}
\usepackage{amssymb}
\usepackage{latexsym}
\usepackage{graphicx}
\usepackage{color}
\usepackage{subcaption}
\usepackage{mathrsfs}
\usepackage{slashed}
\usepackage{soul}
\usepackage{array}
\usepackage{cite}
\usepackage{placeins}
\usepackage{hyperref}

\usepackage{enumerate}
\usepackage{multirow}
\usepackage[table]{xcolor}
\usepackage{colortbl}
\usepackage[utf8]{inputenc}

\usepackage{float}

\usepackage{axodraw4j}
\usepackage[text={7in,10in}]{geometry}

\usepackage{comment}
\includecomment{toexclude} % you can name the comment as you wish
\excludecomment{addsections} % you can name the comment as you wish

\numberwithin{equation}{section}

 \def\be   {\begin{equation}}   \def\ee   {\end{equation}}
       \def\ea   {\end{array}}
 \def\bea  {\begin{eqnarray}}   \def\eea  {\end{eqnarray}}
 \def\bean {\begin{eqnarray*}}  \def\eean {\end{eqnarray*}}

\begin{document}
\hfill\textit{DO-TH 18/13}

%\vspace{7cm}

\begin{center}
%\vspace{5cm}
%\def\thefootnote{\fnsymbol{footnote}}
{\Huge
Dark Matter from Freeze-In via the Neutrino Portal
}
\\ [2.5cm]
{\large{\textsc{ 
Mathias Becker\footnote{\textsl{mathias.becker@tu-dortmund.de}}
}}}
\\[1cm]

\large{\textit{
Fakult\"at f\"ur Physik, Technische Universit\"at Dortmund,\\
44221 Dortmund, Germany
}}
\\ [2 cm]
{ \large{\textrm{
Abstract
}}}
\\ [1.5cm]
\end{center}

We investigate a minimal neutrino portal dark matter (DM) model where a right-handed neutrino both generates the observed neutrino masses and mediates between the SM and the dark sector, which consists of a fermion and a boson. In contrast to earlier work, we explore regions of the parameter space where DM is produced via freeze-in instead of freeze-out motivated by the small neutrino Yukawa couplings in case of $\mathcal{O} \left( \mathrm{TeV} \right)$ heavy neutrinos. \\
For a non-resonant production of DM, its energy density is independent of the DM mass. Assuming a democratic coupling structure we find $M_N \approx 10 \, \mathrm{TeV}$. For the resonant production of DM, we find that it can be produced via freeze-in or freeze-out even with couplings of $\mathcal{O} \left( 10^{-5} \right)$. However, the measurement of the Lyman-$\alpha$ forest rules out the feeble coupled freeze-out case completely, while the resonant freeze-in production is only viable for $m_{DM} \gtrsim 3 \, \mathring{keV}$.

\def\thefootnote{\arabic{footnote}}
\setcounter{footnote}{0}
\pagestyle{empty}

\newpage
\pagestyle{plain}
\setcounter{page}{1}

\section{Introduction} \label{ch:intro}
\begin{figure}[H]
	\begin{center}
		\begin{minipage}{0.48\textwidth}
		\includegraphics[width=80mm]{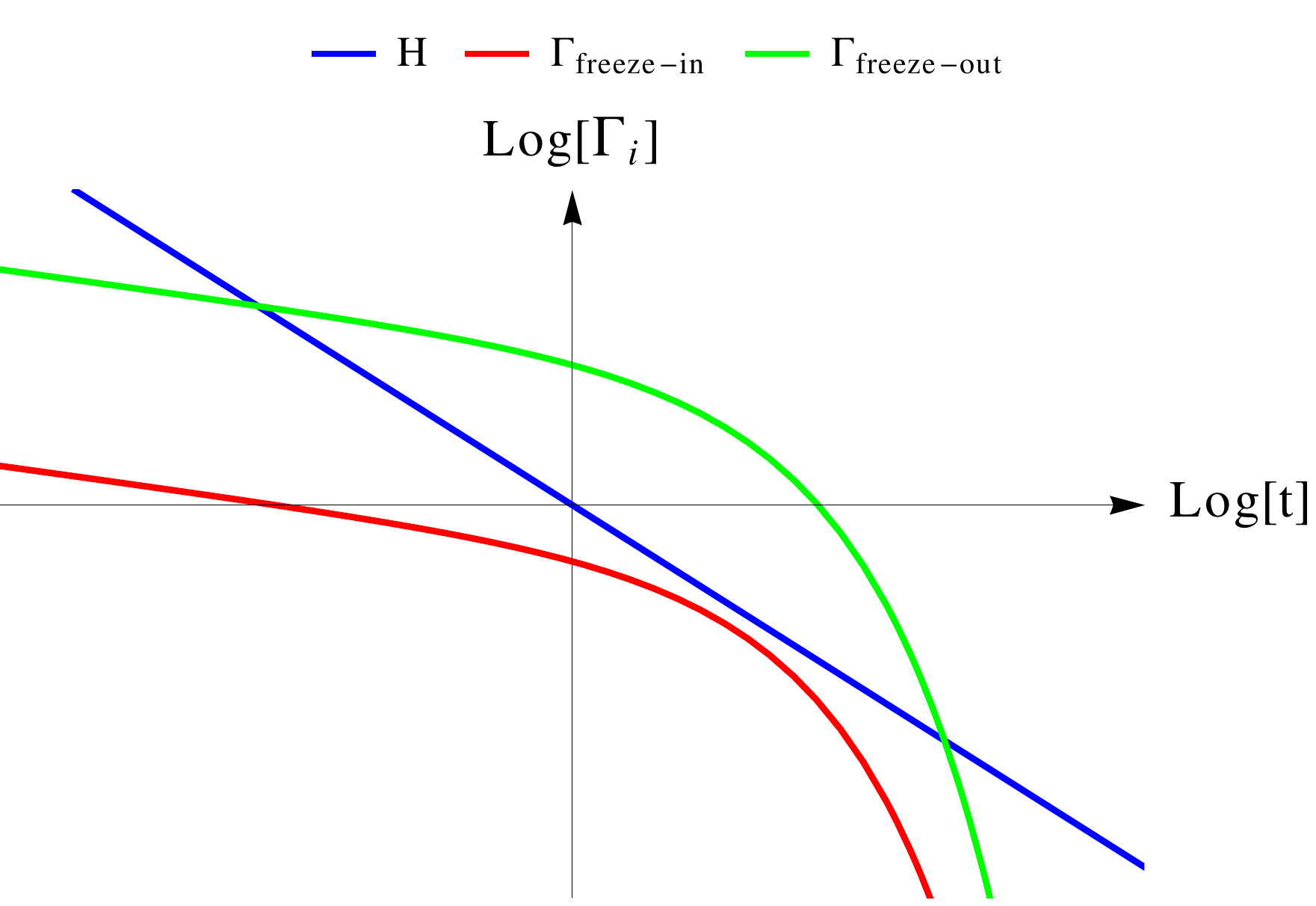}
	\end{minipage}
	\begin{minipage}{0.48\textwidth}
		\includegraphics[width=80mm]{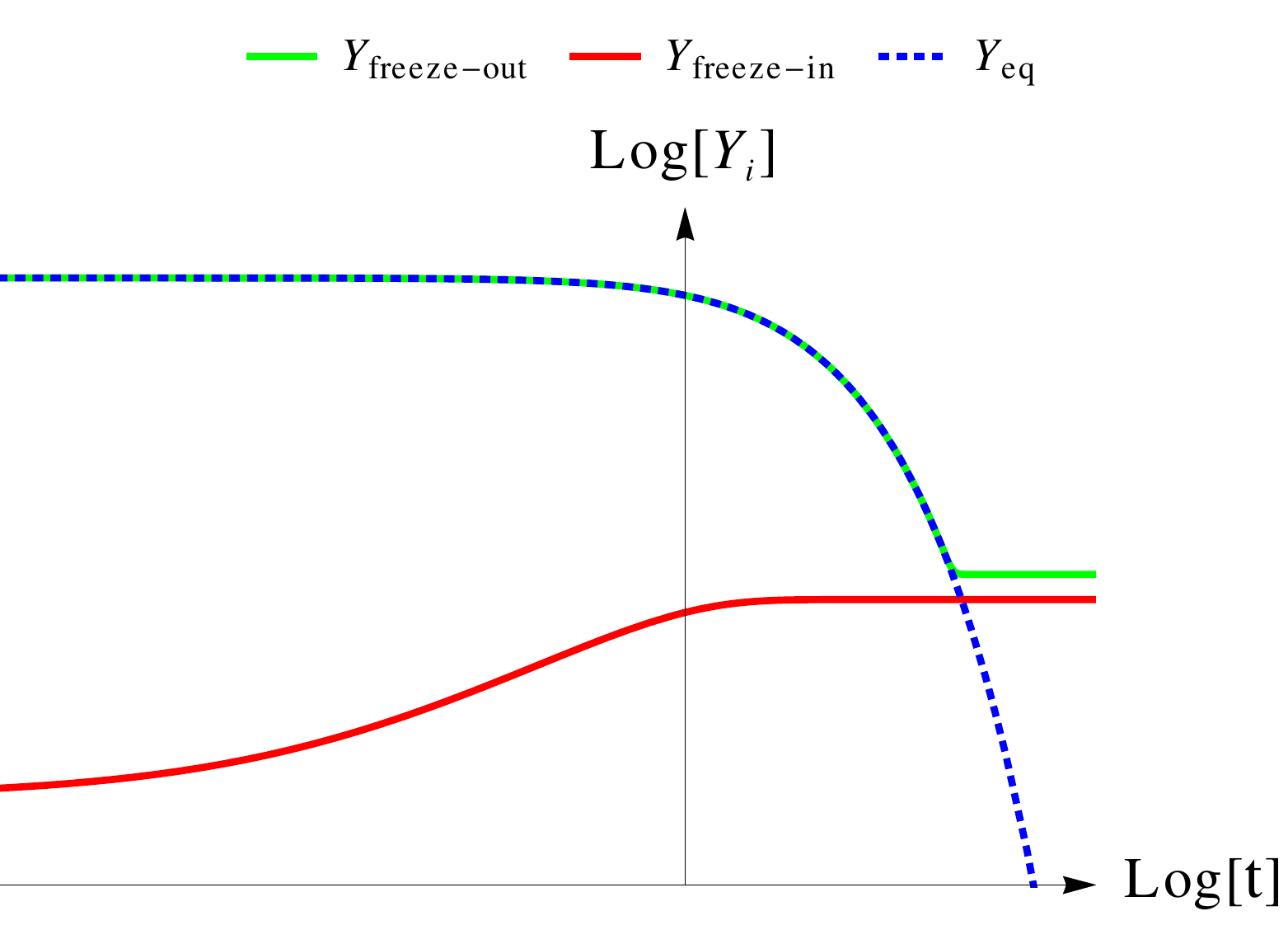}
	\end{minipage}
\end{center}
	\caption{\textbf{Freeze-in and freeze-out scenarios in comparison}:
	 The left panel compares two interaction rates to the Hubble parameter $H$. Both of them are smaller than $H$ for large temperatures since $\Gamma \sim T$ for $T \gg M$ and $H \sim T^2 M_\text{Pl}^{-1}$ and both interaction rates are exponentially suppressed for temperatures $T \sim M$, where $M$ is the DM mass. The difference between the freeze-out case (green) and the freeze-in case (red) results from the much smaller coupling in the freeze-in case. The right panel shows the corresponding number densities compared to the equilibrium density in a co-moving volume. }
	\label{fig:example}
\end{figure}

Both Dark Matter (DM) and neutrino masses provide strong hints for beyond standard model physics (BSM).
%The most popular approach to explain the observed DM density is the WIMP paradigm, where a TeV scale particle can account for the correct DM density via the freeze-out mechanism in case of interactions which are as strong as the weak interaction. \\
%Another evidence for BSM physics is the observation of neutrino oscillations, which require neutrinos to be massive. 
A way to accommodate neutrino masses is to introduce right-handed neutrinos as SM singlets, thereby allowing for mass generation via the type I seesaw mechanism. \\
Furthermore, the resulting heavy neutrino state $N$ is massive and electrically neutral. If it is considered to be a DM candidate it must be stable. Thus, its mass must satisfy $M_N < 2 m_e$. Therefore, the Yukawa coupling has to be very small, namely $y_\nu \lesssim 10^{-6}$. Consequently, the production rate is small, allowing for DM production via the freeze-in mechanism \footnote{A small DM production rate could also be generated by a large mediator mass as was pointed out in \cite{Chen:2017kvz}.} \cite{Hall:2009bx,Bernal:2017kxu}.  \\
In freeze-in scenarios, DM production never becomes efficient, i.e. the interaction rate $\Gamma$ is always small compared to the Hubble parameter $H$, $\Gamma \lesssim H$ (see figure \ref{fig:example}).
To account for the observed DM relic abundance via freeze-in of the decay $h \rightarrow N \nu$, the heavy neutrino mass should be of $\mathcal{O} \left( 10 \, \mathrm{keV} \right)$. However, the possibility of keV sterile neutrino DM via freeze-in within a minimal setup, the Dodelson-Wodrow mechanism\cite{Dodelson:1993je},  is already excluded by the experiment, more precisely by the non-observation of the decay $N \rightarrow \nu \gamma$ \cite{Perez:2016tcq,Campos:2017odj} and Lyman-$\alpha$ measurements \cite{Asaka:2006nq}. However, the idea of sterile neutrino dark matter via different production mechanisms continues to be widely discussed \cite{Adhikari:2016bei}. \\
In case of $M_N > 2 m_e$, the heavy neutrino $N$ is obviously not stable and therefore not a DM candidate. But even in this case the right-handed neutrino can act as a mediator to DM since it is a SM singlet, a possibility which is referred to as \textit{neutrino portal DM} (NPDM)\cite{Escudero:2016tzx,Escudero:2016ksa,Folgado:2018qlv,Batell:2017cmf}. \\
Within these works, the small neutrino masses are generated by the type I seesaw mechanism and DM is produced via the freeze-out mechanism. In contrast, this work explores a minimal NPDM model where DM is produced via the freeze-in mechanism. \\
In section \ref{ch:Setup}, we introduce the particle content and the coupling structure of the model. In section \ref{ch:Boltzmann} the method for deriving the analytic results for the DM number density while assuming a thermal shape of the distribution function is introduced. Although those analytic results, which are discussed in \ref{ch:Analytic}, are not exact they allow for studying the parametrics for DM production. Following in chapter \ref{ch:Numerical} we numerically solve the Boltzmann equations at the level of momentum distribution functions taking the non-thermal form of the momentum distribution into account. Chapter \ref{ch:Bounds} summarizes the relevant constraints on the model from direct detection, lepton flavour violation and structure formation. After that we conclude. \\
Within the appendices, the relevant reduced cross sections are given and the method for solving the boltzmann equations at the level of momentum distribution functions is discussed in more detail.   
\section{Setup}\label{ch:Setup}

A model with similar particle content was investigated in \cite{Escudero:2016ksa}, where DM production within freeze-out scenarios was explored. In addition to the SM particle content, the model includes three right-handed neutrinos $\nu_{R_i}$ to accommodate the observed neutrino masses. The dark sector consists of a fermion $\chi$ and a scalar $\phi$. While they are uncharged under the SM gauge groups, they are charged under a dark symmetry, e.g. a dark U(1) or a $\mathcal{Z}_2$. Assuming the SM particles to be uncharged under the dark symmetry renders the lighter particle of $\chi$ and $\phi$ to be a stable DM candidate since the dark symmetry forbids couplings between SM and dark sector particles. 
In this scenario, the resulting heavy neutrinos $N_i$ mediate between the DM and the SM particles since the singlets $\nu_{R_i}$ can couple to $\bar{\chi}$ and $\phi$ via a Yukawa coupling as long as the expression $\bar{\chi} \phi$ is a singlet under all gauge groups. The parts of the Lagrangian relevant for the neutrino mass generation and the coupling to DM are given by
\begin{align}
 \mathcal{L} \supset - \underbrace{ \left( Y_\nu \right)_{ij} \bar{\nu}_{L_i} h \nu_{R_j} - \frac{1}{2} \left( M_M \right)_{ij} \bar{\nu}_{R_i}^C \nu_{R_j}}_\text{Neutrino mass generation} - \underbrace{y_\chi \phi \bar{\chi} \nu_{R_i}}_\text{DM coupling} + h.c.    \quad .
\end{align}
Here, we assumed a universal coupling of DM to the three right-handed neutrinos.
Furthermore, we do not take into account any contribution to the DM relic abundance from a possible Higgs portal interaction arising from the term $\left(\phi \phi^* \right) \left( h h^* \right)$ in the scalar potential and additionally assume that $\phi$ does not acquire a VEV \footnote{In fact, the validity of this assumption as well as the vacuum stability of this model will be investigated in a future work since due to a fermion loop consisting of a $\nu_R$ and a $\chi$ the $\phi$ mass term receives a negative contribution. In case the fermions in the loop are heavy compared to the boson those radiative corrections might lead to a negative $m_\phi^2$ and thereby break the symmetry that stabilizes DM. Similar effects have been investigated for the scotogenic model \cite{Merle:2015gea,Lindner:2016kqk} where those effects constrain the parameter space significantly. }. Moreover, effects resulting from kinetic mixing of possible vector mediators of the dark symmetry with the SM gauge bosons are neglected. Thus, our analysis focuses on the neutrino portal to DM only. \\
After electroweak symmetry breaking the observed light neutrino masses are generated via the type I seesaw mechanism. To ensure that the observed neutrino masses and mixing angles are reproduced we utilize the following parametrization of the Yukawa coupling matrix $Y_\nu$ \cite{Casas:2001sr}:
\begin{align}
 Y_\nu = \frac{\sqrt{M_N}}{v} R \sqrt{m_\nu} U_\text{PMNS}^\dagger = \underbrace{\frac{\sqrt{M_N \Delta m_\nu}}{v}}_{\equiv y_\nu} \underbrace{R \frac{1}{\sqrt{\Delta m_\nu}} \sqrt{m_\nu}}_{\equiv R'} U_\text{PMNS}^\dagger \, , 
\end{align}
where we assumed the Majorana mass matrix $M_M$ to be diagonal with degenerated eigenvalues, i.e. $M_M = \text{diag} \left( M_N, M_N, M_N \right)$. $U_\text{PMNS}$ is the PMNS matrix, $v$ is the vacuum expectation value of the Higgs field, $\sqrt{m_\nu}$ is a diagonal matrix with the square root of the neutrino masses as eigenvalues, $R$ is an orthogonal complex $3 \times 3$ matrix and $\Delta m_\nu$ is the square root of the large mass squared difference $\Delta m_\nu = \sqrt{\Delta m_\nu^2}$. The mass- and interaction eigenstates are transformed into each other in leading order in the small parameter $y_\nu v M_N^{-1}$ by the matrix $U$ : 
\begin{align}
 \begin{pmatrix}
  \nu_{L} \\
  \nu_R
 \end{pmatrix} = U \begin{pmatrix}
  \nu \\
  N \end{pmatrix}
  \approx
  \begin{pmatrix}
  U_\text{PMNS} & Y_\nu^T v M_N^{-1} \\
  -Y_\nu U_\text{PMNS} \, v M_N^{-1} & 1
 \end{pmatrix}
 \begin{pmatrix}
  \nu \\
  N \end{pmatrix}
\end{align}

%They are described in terms of the interaction eigenstates by:
%\begin{align}
%\begin{pmatrix}
% \nu \\
% N
%\end{pmatrix} = U 
%\begin{pmatrix}
% \nu_L \\
% \nu_R
%\end{pmatrix} \approx 
%\begin{pmatrix}
% \sqrt{1 - \frac{y^2 v^2}{M_N^2}} & -\frac{y v}{M_N} \\
% \frac{y v}{M_N} & \sqrt{1 - \frac{y^2 v^2}{M_N^2}}
%\end{pmatrix}
%\begin{pmatrix}
% \nu_L \\
% \nu_R
%\end{pmatrix} \, .
%\end{align}
The mixing between the left and right handed neutrinos causes an interaction between $\nu$, $N$ and the Higgs as well as a coupling of $N$ to the $SU(2)_L$ gauge bosons. As presented in \cite{Pilaftsis:1991ug}, the resulting interactions between the heavy and the light neutrinos are given by:
\begin{align}
 \mathcal{L}_W \supset &- \frac{g_W}{2 \sqrt{2}} l_i W^-_\mu \gamma^\mu \left( 1 - \gamma_5 \right) B_{l_i N_j} N_j + h.c. \, , \\
 \mathcal{L}_Z \supset &- \frac{g_W}{4 \cos \left( \Theta_W \right)} Z^0_\mu \left\lbrace \bar{\nu_i} \gamma^\mu \left[ i \text{Im} \left( C_{\nu_i N_j} \right) - \gamma_5 \text{Re} \left( C_{\nu_i N_j} \right) \right] N_j \right. \\
 &\left. \bar{N_i} \gamma^\mu \left[ i \text{Im} \left( C_{N_i N_j} \right) - \gamma_5 \text{Re} \left( C_{N_i N_j} \right) \right] N_j   + h.c.  \right\rbrace \nonumber \, , \\
 \mathcal{L}_H \supset &-\frac{g_W}{4 M_W} h \left\lbrace 2 \bar{\nu_i} \left[ \left( m_{\nu_i} + M_{N_j} \right) \text{Re} \left( C_{\nu_i N_j} \right) + i \gamma_5 \left( M_{N_j} - m_{\nu_j} \right) \text{Im} \left( C_{\nu_i N_j} \right)  \right] N_j \right. \\
 &\left. + \bar{N_i} \left( M_{N_i} + M_{N_j} \right) \text{Re} \left( C_{N_i N_j} \right)  N_j \right\rbrace   \, . \nonumber
\end{align}
The matrices $B$ and $C$ are defined as in \cite{Pilaftsis:1991ug} and in case of real Yukawa couplings, as we will assume no CP violation from now on, they yield:
\begin{align}
B_{l_i N_j} \approx \frac{v}{M_N} \left( Y_\nu^T \right)_{ij} \, , \quad 
C_{\nu_i N_j} \approx \frac{v}{M_N} \left( U_\text{PMNS}^T Y_\nu^T \right)_{ij} \, , \quad C_{N_i N_j} \approx \frac{v^2}{M_N^2} \left( Y_\nu Y_\nu^T \right)_{ij} \, .
\end{align}
Thus, the couplings relevant for heavy neutrino production are given by
\begin{align}
 \mathcal{L}_W &\supset - \frac{M_W y_\nu}{\sqrt{2} M_N} \left( U_\text{PMNS} R'^T \right)_{ij} \bar{l_i} W^-_\mu \gamma^\mu \left( 1 - \gamma_5 \right)  N_j + h.c \label{eq:couplingW}. \, , \\
 \mathcal{L}_Z &\supset    \frac{M_W y_\nu}{2 \cos \left( \Theta_W \right) M_N} \left( R'^T \right)_{ij} Z^0_\mu \bar{\nu_i} \gamma^\mu  \gamma_5 N_j \label{eq:couplingZ} \, , \\
 \mathcal{L}_H &\supset -y_\nu h \left( R'^T \right)_{ij} \bar{\nu}_i N_j - y_\nu^2 \frac{v}{M_N} h \left( R'^T R' \right)_{ij} \bar{N_i} N_j \label{eq:couplingH} \, ,
\end{align}
whereas the coupling of the heavy neutrino to the dark sector is governed by:
\begin{align}
\mathcal{L}_\chi \supset - y_\chi \phi \bar{\chi} N_i + h.c \quad .
\end{align}
Note that the parameters $y_\nu$ and $M_N$ are not independent and related by the seesaw mechanism requiring $y_\nu = \sqrt{\Delta m_\nu M_N }v^{-1}$. Therefore, the couplings in eq. \eqref{eq:couplingW}-\eqref{eq:couplingH} excluding the flavor dependent part can be rewritten as:
\begin{align}
\begin{matrix}
g_{h\nu N} = y_\nu = \frac{\sqrt{m_\nu M_N}}{v} & g_{WlN,Z\nu N} = y_\nu \frac{M_W}{M_N} = \sqrt{\frac{m_\nu}{M_N}} \frac{M_W}{v} \\ 
 g_{hNN} = y_\nu^2 \frac{v}{M_N} = \frac{m_\nu}{v} & g_{ZNN} = g_{Z \nu N} \frac{y_\nu v}{M_N} = \frac{m_\nu}{M_N} \frac{M_W}{v} 
\end{matrix}
\end{align}
Thus, for $M_N \geq M_W$, the coupling $g_{h \nu N}$ can be expected to be dominant and the $h \nu N$ vertex is the most relevant one for DM production. Whereas for $M_N \leq M_W$, the $W l N$ and $Z \nu N$ vertices are expected to contribute the most to DM production as long as $M_N \gtrsim m_\nu$. 

\section{Boltzmann Equations} \label{ch:Boltzmann}
Determining the relic abundance of the DM candidate requires solving the Boltzmann equations, which describe the time evolution of the particle number densities in the expanding universe. In principle, the boltzmann equations have to be solved at the level of momentum distribution functions, which then are integrated to obtain the number density. For a freeze-out production of DM however those distribution functions can be safely assumed to be proportional to a Boltzmann distribution, which allows for solving the Boltzmann equations at the level of number densities directly. Although this assumption can lead to less precise results in case of freeze-in production we will still use this formalism to obtain analytic expressions for the relic density in chapter \ref{ch:Analytic}. Later on in chapter \ref{ch:Numerical}, a numerical solution of the Boltzmann equation is given at the level of momentum distribution functions. \\
Here, we review the formalism for solving the Boltzmann equation for number densities, while the one for distribution functions is discussed in appendix \ref{app:boltzmann}. \\
Adopting the formalism used in \cite{Giudice:2003jh}, the Boltzmann equations can be written as
\begin{align}
\dot{n}_N + 3 H n_N = - \sum_{a,i,j,\dots} \left( \frac{n_N n_a \dots}{n_N^\text{eq} n_a^\text{eq} \dots} \gamma_\text{eq} \left( Na \dots \rightarrow ij\dots \right) - \frac{n_i n_j \dots}{n_i^\text{eq} n_j^\text{eq} \dots} \gamma_\text{eq} \left( ij \dots \rightarrow Na\dots \right) \right) \, .
\end{align}
Here, $n_i$ is the number density of particle species $i$. The $3 H n_N$ term takes the expansion of the universe into account while the right hand side governs the impact of scattering processes which occur with a certain thermal rate $\gamma_\text{eq}$. 
The equilibrium number densities $n_i^\text{eq}$ are given by the momentum integral over the distribution function $f_i^\text{eq}$ of the respective particle species which is approximated with a Boltzmann distribution in our case:
\begin{align}
n_i^\text{eq} = \int \frac{d^3 p}{\left( 2 \pi \right)^3} f_i^\text{eq} = \frac{g_i}{2 \pi^2} m_i^2 T K_2 \left( \frac{m_i}{T} \right) \, . \label{eq:eqdens}
\end{align}
For a two to two scattering involving only CP conserving interactions the quantity $\gamma_\text{eq}$ results in
\begin{align}
\gamma_\text{eq} \left( Na  \rightarrow ij \right)  = \gamma_\text{eq} \left( ij  \rightarrow Na \right) = \frac{T}{64 \pi^4} \int_{s_\text{min}}^\infty ds \sqrt{s} \hat{\sigma} \left( s \right) K_1 \left( \frac{\sqrt{s}}{T} \right) \, , \label{eq:thrate22}
\end{align}
where $\hat{\sigma} \left( s \right) = 2 s \, \sigma \left( s \right) \lambda\left[ 1 , \frac{m_N^2}{s}, \frac{m_a^2}{s} \right]$ with $\lambda \left[ a ,b ,c \right] = \left( a - b -c \right)^2 - 4 bc$, $K_1 \left( x \right)$ is a Bessel function and $s_\text{min} =\text{max} \left[ \left( m_a + M_N \right)^2 , \left( m_i +m_j \right)^2 \right]$. \\
Next, to simplify the form of the Boltzmann equations we write them in terms of the quantity $Y=\frac{n}{s_E}$, where $s_E = \frac{2 \pi^2 g_\text{eff}^s}{45} T^3$ is the entropy density. This leads to 
\begin{align}
z H s_E \frac{dY_N}{dz} = - \sum_{a,i,j,\dots} \gamma_\text{eq} \left( Na \dots \leftrightarrow ij\dots \right) \left[ \frac{n_N n_a \dots}{n_N^\text{eq} n_a^\text{eq} \dots}  - \frac{n_i n_j \dots}{n_i^\text{eq} n_j^\text{eq} \dots} \right]\, , \label{eq:boltzmanneq}
\end{align}
with $z=\frac{M_N}{T}$. \\
For the special case of freeze-in production via a two-to-two scattering process $b_1 b_2 \rightarrow i j$ the solution of this equation can be written in a compact form. Here, $b_{1/2}$ are particles in thermal equilibrium with the SM, whereas the number densities of $i$ and $j$ satisfy $n_{i/j} \ll n_{i/j}^\text{eq}$. Then, the Boltzmann equation for the particle species $i$ is given by:
\begin{align}
 z H s_E \frac{dY_{i}}{dz} = \gamma_{eq} \left( b_1 b_2 \leftrightarrow i j  \right).  
\end{align}
Inserting $\gamma_{eq}$ \eqref{eq:thrate22} and integrating the equation from very large temperatures, i.e. $z \rightarrow 0$, up to today, i.e. $z \rightarrow \infty$, yields:
\begin{align}
 Y_i = \frac{1}{ 64 K m_i^4 \pi^4} \int \limits_0^{\infty} dz \, z^3 \int \limits_{s_\text{min}}^\infty ds \, \sqrt{s} \hat{\sigma} \left( s \right) K_1 \left( \frac{\sqrt{s}}{m_i} z \right) \, . \label{eq:densitycalc}
\end{align}
Here we use $K = H s_E T^{-5}$ and $z= m_i T^{-1}$. After performing the $z$ integration with the initial condition $Y_i \left( z=0 \right) = 0$ we are left with\footnote{Eq. \eqref{eq:niceformula} illustrates a behaviour typical for the freeze-in mechanism: Assuming the reaction $ b_1 b_2 \leftrightarrow i j $ involves a dominant mass scale $M_\text{max}$ and noting that the mass dimension of the remaining integral is minus one yields $Y_i \sim M_\text{max}^{-1}$.}
\begin{align}
Y_i = \frac{3}{128 K \pi^3} \int \limits_{s_\text{min}}^\infty ds \, \frac{\hat{\sigma} \left( s \right)}{\sqrt{s^3}} \, . \label{eq:niceformula}
\end{align}

%%%%%%%%%%%%%%%%%%%%%%%%%%%%%%%%%%%%%
\section{Relic Abundance: Analytic Estimates} \label{ch:Analytic}
\begin{figure}
%		\begin{subfigure}[c]{0.95\textwidth}
			\begin{center}
	\includegraphics[width=9cm]{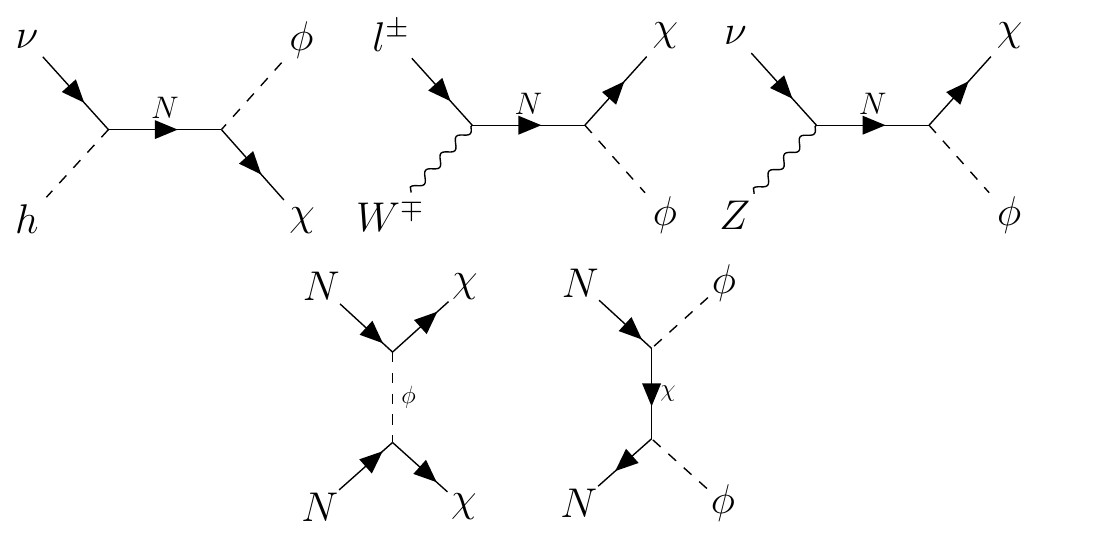}
% \feynmandiagram [scale=0.75,transform shape][horizontal=a to b] {
% 	i1 [particle=\(\nu\)] -- [fermion] a -- [scalar] i2 [particle=\(h\)],
% 	a -- [fermion, edge label=\(N\)] b,
% 	f2[particle=\(\phi\)] -- [scalar] b -- [fermion] f1 [particle=\(\chi\)],
% }; 
% \feynmandiagram[scale=0.75,transform shape][horizontal=f2 to f3] {
% 	f1 [particle=\(l^\pm\)] -- [fermion] f2 -- [fermion, edge label=\(N\)] f3 -- [fermion] f4 [particle=\(\chi\)],
% 	f2 -- [boson] p1 [particle=\(W^\mp\)],
% 	f3 -- [scalar] p2 [particle=\(\phi\)],
% };
%   \feynmandiagram[scale=0.75,transform shape][horizontal=f2 to f3] {
%   	f1 [particle=\(\nu\)] -- [fermion] f2 -- [fermion, edge label=\(N\)] f3 -- [fermion] f4 [particle=\(\chi\)],
%   	f2 -- [boson] p1 [particle=\(Z\)],
%   	f3 -- [scalar] p2 [particle=\(\phi\)],
%   }; 
%   \end{center}
%\end{subfigure}  
% \begin{subfigure}[c]{0.95\textwidth}
% 	\begin{center}
%      \feynmandiagram [scale=0.6,transform shape][vertical=a to b] {
%   	 	i1 [particle=\(\chi\)]  -- [anti fermion]  a  -- [ anti fermion]  i2 [particle=\(N\)],
%   	 	a -- [scalar, edge label=\(\phi\)] b,
%   	 	f1 [particle=\(\chi\)] -- [anti fermion]  b -- [anti fermion] f2[particle=\(N\)] ,
%   	 }; \quad
%   	  \feynmandiagram [scale=0.6,transform shape][vertical=a to b] {
%   	  	i1 [particle=\(\phi\)]  -- [scalar]  a  -- [anti fermion]  i2 [particle=\(N\)],
%   	  	a -- [fermion, edge label=\(\chi\)] b,
%   	  	f1 [particle=\(\phi\)] -- [scalar]  b -- [fermion] f2[particle=\(N\)] ,
%   	  };
   	  \end{center}
%   	\end{subfigure} 
   \caption{Feynman diagrams for the DM production processes.}
   \label{fig:feynman}
\end{figure}
The $2 \leftrightarrow 2$ scattering processes responsible for producing DM can be classified into two categories: \textit{SM Particle Scattering} and \textit{Heavy Neutrino Scattering}. The SM particle scattering processes involve two SM particles in the initial state, have $\chi$ and $\phi$ in the final state and are mediated by the heavy neutrino. Consequently, we have $\sigma \sim y_\nu^2 y_\chi^2$. \\
The heavy neutrino scattering processes have two heavy neutrinos in the initial state and produce a pair of $\chi$ or $\phi$. Here, we have $\sigma \sim y_\chi^4$.\\
%\begin{itemize}
%\item \textbf{SM Particle Scattering: $\sigma \sim y_\nu^2 y_\chi^2$} \\
%Here, three different processes, all having the heavy neutrino $N$ as the mediator, contribute:
%\begin{enumerate}
%\item $\nu h \leftrightarrow \chi \phi$
%\item $W^\pm l^\mp \leftrightarrow \chi \phi$
%\item $Z \nu \leftrightarrow \chi \phi$
%\end{enumerate}
%As discussed in the end of chapter \ref{ch:Setup}, for $M_N \gg M_W$ only process number one contributes significantly while for $M_N \ll M_W$ we only need to take into account the processes two and three.
%\item \textbf{Heavy Neutrino Scattering: $\sigma \sim y_\chi^4$} \\
%Here, only two processes contribute: One producing the dark scalar $\phi$ and the other one producing the dark fermion $\chi$.
%\begin{enumerate}
%\item $NN \leftrightarrow \chi \chi$
%\item $NN \leftrightarrow \phi \phi$
%\end{enumerate}
%\end{itemize}
All contributing diagrams are displayed in figure \ref{fig:feynman}. The following discussion assumes only one SM and right-handed neutrino generation. However, these results can easily be translated into a three generation setup due to the assumption of degenerated heavy neutrino masses, i.e. $M_{N_i} = M_N$ and the universal coupling of the dark sector to the right-handed neutrinos. For the heavy neutrino scattering, the one generation result has to be multiplied by a factor of nine. For the dominant SM particle scattering process $\nu_i h \rightarrow \chi \phi$ via a $N_j$ the one generation contribution with a neutrino Yukawa coupling of $y_\nu = \sqrt{\Delta m_\nu M_N} v^{-1}$ has to be multiplied by $\sum_i |\sum_j \left( R'^T \right)_{ij}|^2 = f_1 \left( \mathbf{\theta} \right)$ where $\mathbf{\theta}$ is a vector containing the in our case three real angles parametrizing the orthogonal matrix $R$. Choosing the standard parametrization for an orthogonal three by three matrix we find $10^{-16} \lesssim f_1 \left( \theta \right) \leq 3$. \\
Since the $Z \nu_i N_j$ vertex has the same flavor structure as the $h \nu_i N_j$ vertex the one generation result for the $Z \nu$ initial state is multiplied by the same factor as the $h \nu$ initial state. \\
Only for the $W l$ initial the factor differs and results in $f_2 \left( \theta \right) = \sum_i |\sum_j \left( U_\text{PMNS} R'^T \right)_{ij}|^2 $ . Here, we find $10^{-18} \lesssim f_2 \left( \theta \right) \lesssim 7.65 $. Scanning both $f_1$ and $f_2$ for randomly chosen values for the angles $\theta$ shows that on average $f_2 \approx 2.5 f_1$. Nevertheless, excluding the cases where $f_1$ is close to its lower bound, the contribution of the $h \nu_i$ initial state is still the dominant one due to the following reason: The production via the scattering of the gauge bosons is only viable for temperatures below the critical temperature where the $SU(2)_L \times U(1)_Y$ symmetry of the SM gets broken. Hence, the time of production is small compared to the Higgs neutrino scattering. Therefore, we consider only the production via $h \nu_i \rightarrow \chi \phi$ for the analytic estimates, while all production channels are taken into account in the numerical solution. 
\subsection{SM Particle Scattering}
For the rest of the discussion, we assume that the dark sector particles have roughly the same mass and replace $m_\phi = m_\chi$. 
%As discussed in the end of chapter \ref{ch:Setup}, for $M_N \gg M_W$ the coupling of the heavy neutrino to the Higgs and a light neutrino is much larger than the coupling to the $SU(2)$ gauge bosons.
The reduced cross section for the dominant production channel is given by:
\begin{align}
 \sigma_{vh \leftrightarrow \chi \phi } \left( s \right) = y_\chi^2 y_\nu^2 \frac{ \left( 1 - \frac{m_h^2}{s} \right) s^2 \sqrt{ \left( 1 - 4 \frac{m_\chi^2}{s} \right)}}{16 \pi \left[ \left( s - M_N^2 \right)^2 + \Gamma_N^2 M_N^2 \right]} \, . \label{eq:SMScatteringapprox}
\end{align}  
Here, $\Gamma_N$ is the total decay width of the propagating neutrino. There are two cases to be distinguished:
\begin{itemize}
\item The resonant case with $M_N \geq 2 m_\chi$ where $M_N^2 \geq s_\text{min}$.
\item The non-resonant case with $M_N < 2 m_\chi $ where $M_N^2 < s_\text{min}$.
\end{itemize} 
First, we discuss the non-resonant case.
If we neglect the contribution of the Higgs mass, i.e. $m_h \ll m_\chi$, we can use eq. \eqref{eq:niceformula} to determine the relic density directly:
\begin{align}
Y_\text{DM}= Y_\chi + Y_\phi &= \frac{3^4}{2^{11} \pi^5} \frac{y_\nu^2 y_\chi^2}{\sqrt{g_\text{eff}}g_\text{eff}^s} \frac{M_\text{pl}}{\sqrt[4]{\left[ 4 m_\chi^2 - M_N^2 \right]^2 + \Gamma_N^2 M_N^2} } \label{eq:resultnr} \\
&\overset{M_N \ll m_\chi}{=} \frac{3^4}{2^{12} \pi^5} \frac{y_\nu^2 y_\chi^2}{\sqrt{g_\text{eff}}g_\text{eff}^s} \frac{M_\text{pl}}{m_\chi} \label{eq:resultnr2}  \, ,
\end{align}
where $g_\text{eff}^{(s)}$ are the number of effective relativistic (entropy) degrees of freedom which are both assumed to be constant during this calculation with $g_\text{eff}^{(s)} = 106.75$\footnote{This is a good approximation as long as the production is mainly efficient for temperatures above $100 \, \mathrm{GeV}$.}. Note that for obtaining this result the reduced cross section was multiplied by an additional factor of four arising from the four degrees of freedom of the Higgs doublet before the electroweak phase transition. \\
Remarkably in case of a heavy DM mass $m_\chi$ compared to the mediator mass $M_N$, the result is inversely proportional to the DM mass, i.e. the energy density is independent of $m_\chi$. This allows for predicting the value of the product of the Yukawa couplings $y_\nu y_\chi$ by setting $Y_{DM} \left( z \rightarrow \infty \right) = Y_{DM,\text{exp}}$, with
\begin{align}
Y_{DM,\text{exp}} = \frac{\Omega_{DM}}{\Omega_{B}} \frac{m_B}{m_{DM}} Y_B \approx 10^{-10} m_B m_\chi^{-1} \label{eq:DMexp} \, .
\end{align}
The experimental values for $\Omega_{DM}$, the density parameter for baryons $\Omega_B$, and the baryon number density in a co-moving volume $Y_B$, are taken from \cite{Patrignani:2016xqp} and $m_B$, the average baryon mass, is approximated with the proton mass. \\
Evaluating $Y_{DM} = Y_{DM,\text{exp}}$ results in:
\begin{align}
\left( y_\nu y_\chi \right)^2 \approx 10^{-3} \frac{m_B}{M_\text{Pl}} \approx 10^{-21} \, .
\end{align}
The implications of this result are discussed in chapter \ref{ch:DisAna} \\
%Before we proceed with the discussion of the resonant production we want to comment on the validity of the approximation of $m_h = 0$ taken in order to arrive at the result in eq. \eqref{eq:resultnr}. In princepale, the $z$ integral in eq. \eqref{eq:densitycalc} has to be split into two parts, namely from $z=0$ to $z \left( T_c \right)$ with $m_h \neq 0$ and from $z \left( T_c \right)$ to $z \rightarrow \infty$ with $m_h = 0$ where $T_c$ is the critical temperature where the $SU(2) \times U(1)$ symmetry is broken and the higgs and the gauge bosons become massive. For $m_\chi > \frac{1}{2} m_h$ and $m_\chi \gg M_N$ the integration is still solvable analytically and shows that the deviation of the approximated result \eqref{eq:resultnr2} is $6 \%$ at $m_\chi = \frac{m_h}{2}$ and goes to zero for larger $m_\chi$. Thus, the result is trustworthy as long as $m_\chi > \frac{m_h}{2}$. \\  
Next, we discuss the resonant case, i.e. $M_N \geq 2 m_\chi$.
As it was pointed out in \cite{Blennow:2013jba}, in this case it is useful to approximate the Breit-Wigner peak in eq. \eqref{eq:SMScatteringapprox} with:
\begin{align}
\int_c^{\infty} dx \, \frac{f\left( x \right)}{\left( x - a \right)^2 + b^2} \approx \frac{f \left( a \right)}{b} \, ,
\end{align}
which is valid as long as $b \ll a$, i.e. $\Gamma_N \ll M_N$.
Then, the integration of eq. \eqref{eq:niceformula} results in:
\begin{align}
 Y_{DM} \left( z \rightarrow \infty \right) = \frac{27}{4 \pi^5 g_\text{eff}^s \sqrt{g_\text{eff}}} \frac{\left( y_\nu y_\chi \right)^2}{y_\nu^2 + y_\chi^2} \frac{M_\text{pl}}{M_N} \, ,
\end{align}
where we already used $M_N \gg m_\chi$ to simplify the result. 
%Since the result is not proportional to $m_\chi^{-1}$ fitting the observed DM density depends on the DM mass $m_\chi$ itself. 
Again, we postpone the discussion of the result to chapter \ref{ch:DisAna}. 

\subsection{Heavy Neutrino Scattering}
The cross sections for the heavy neutrino scattering for the case of $M_N \ll m_\chi$ result in
\begin{align}
 \sigma_{NN \rightarrow \chi \chi} &=y_\chi^4 \frac{\sqrt{1 - \frac{4 m_\chi^2}{s}}}{8 \pi s} \, , \\
 \sigma_{NN \rightarrow \phi \phi} &=\frac{y_\chi^4}{2 \pi} \left[ \left( 1 + 4 \frac{m_\chi^2}{s} \right) \log \left( \frac{s-2 m_\chi^2 - \sqrt{s^2 - s m_\chi^2} }{2 m_\chi^2} \right) + 2 \sqrt{1 - 4\frac{m_\chi^2}{s}} \right] \, .
\end{align}
By again employing eq. \eqref{eq:niceformula} we find:
\begin{align}
Y_\text{DM} = Y_\chi + Y_\phi = \frac{35 \cdot 3^3 y_\chi^4}{2^{13} \pi^5 \sqrt{g_\text{eff}} g_\text{eff}^s} \frac{M_\text{Pl}}{m_\chi} \, .
\end{align}
As for the SM particle scattering in the limit of $M_N \ll m_\chi$, the DM density is inversely proportional to its mass. \\ 
%However, for a realistic result we have to consider both SM particle scattering and heavy neutrino scattering processes. This is discussed in the next chapter. \\
For the case where the SM scattering processes are in the resonant regime, i.e. $M_N >2 m_\chi$, in the limit $M_N \gg m_\chi$ we cannot find an analytic estimate for the DM relic density beside 
\begin{align}
Y_\text{DM} \sim \frac{y_\chi^4 M_\text{Pl}}{\sqrt{g_\text{eff}} g_\text{eff}^s M_N} \, .
\end{align} 
Although the factor of proportionality is unknown we expect this to be much smaller compared to the contribution of the SM particle scattering. This is due to the resonance contributing to the production via SM particle scattering. Hence, we neglect this contribution for the discussion of the analytic results.  
\subsection{Discussion of the Analytic Results} \label{ch:DisAna}
\begin{figure}
	\begin{center}
		\includegraphics[width=8cm]{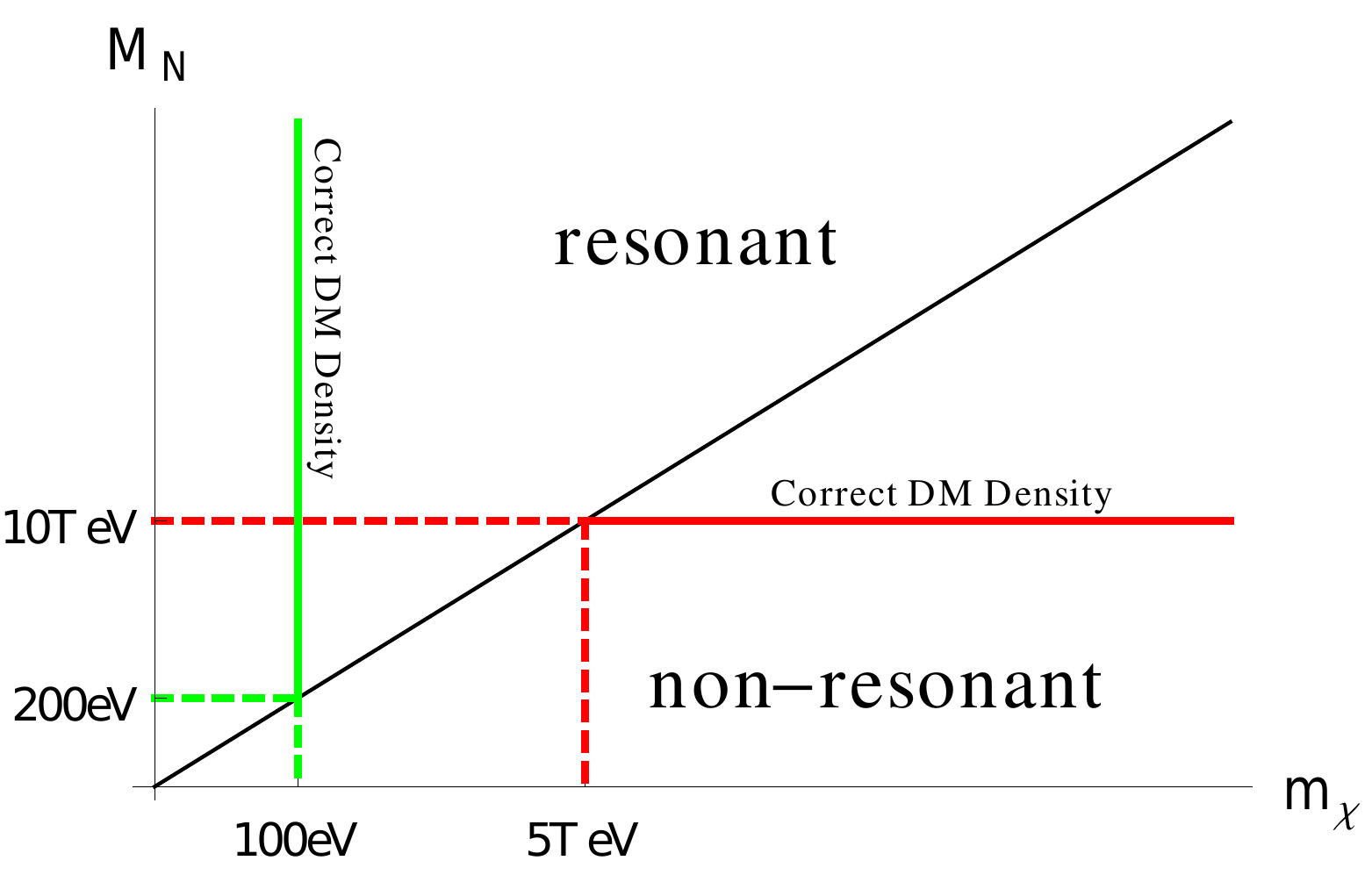}
	\end{center}
	\caption{Parameter space for $y_\chi = y_\nu$: the black line divides the plane spanned by the DM mass $m_\chi$ and the mediator mass $M_N$ into two halves. The upper (lower) half corresponds to the resonant (non-resonant) DM production regime. The red and green line show where the correct amount of DM is produced for the non-resonant and the resonant regime, respectively. 
	In the non-resonant regime, producing the correct density only depends on the mediator mass, whereas it only depends on the DM mass in the resonant region.}
	\label{fig:niceplot}
\end{figure}

In the limit of $M_N \ll m_\chi \approx m_\phi$ we found analytic solutions for the DM relic density for both types of processes. Combining both results yields:
\begin{align}
 Y_{DM} \left( z \rightarrow \infty \right) =  \frac{3^3 }{2^{13} \pi^5 g_\text{eff}^s \sqrt{g_\text{eff}} } \frac{M_\text{Pl}}{m_\chi} \left( 6 y_\nu^2 y_\chi^2 +  35 y_\chi^4 \right) \, . \label{eq:result1}
\end{align}
By comparing this expression with the observed DM density \eqref{eq:DMexp} one obtains
\begin{align}
 \left( 6 y_\nu^2 y_\chi^2 + 35 y_\chi^4 \right) \approx 10^{-21} \, . \label{eq:result2}
\end{align}
Since the coupling $y_\nu$ is only a function of $M_N$ the coupling $y_\chi$ is fixed by the heavy neutrino mass $M_N$. Moreover, we find $y_\chi \lesssim 10^{-5}$ in order not to overproduce DM. \\ In principle, the couplings $y_\chi$ and $y_\nu$ are otherwise unrelated. However, both describe a coupling to the right-handed neutrino and - if the heavy neutrino is lighter than $\mathcal{O} \left( 10^{15} \, \mathrm{GeV} \right)$ - both couplings are required to be relatively small. This motivates the idea that they might be suppressed by the same mechanism, resulting in $y_\nu \approx y_\chi$.\footnote{For example, such a mechanism could be an extra dimensional model where the right-handed neutrino in contrast to all other particles propagates in an extra dimension since it is uncharged under all considered gauge groups. Thereby, its coupling gets suppressed by the reduced wave function overlap \cite{ArkaniHamed:1998vp, Becker:2017ssz}}. Considering a model which generates $y_\chi \approx y_\nu$ allows for constraining the mass of the heavy neutrino since then eq. \eqref{eq:result2} reads
\begin{align}
 41 y_\nu^4 = 41 \left( \frac{m_\nu M_N}{v^2} \right)^2 \approx 10^{-21} \, .
\end{align}
Thus, to fit the observed DM density \eqref{eq:DMexp}, $M_N \approx 10 \, \mathrm{TeV}$ is required. Since we are investigating the non-resonant regime we have $M_N < 2 m_\text{DM}$. Therefore, we find a lower bound on the DM mass of $m_\text{DM} \gtrsim 5 \, \mathrm{TeV}$ if we naively assume the behaviour for large DM masses to be also correct for parameters close to the transition of the non-resonant to resonant regime.  \\
We achieved this result by assuming $n_N = n_N^\text{eq}$, $m_\chi \gg M_N$ and by only taking into account the dominant processes of the SM particle scattering. From eq. \eqref{eq:result1}, we see that the contribution of the heavy neutrino scattering processes accounts for roughly eighty percent of the produced DM in case of $y_\chi = y_\nu$. Thus, the result will be altered significantly if the heavy neutrinos are out of equilibrium during the time where the production via heavy neutrino scattering is efficient. Also, we expect a significant change in areas of the parameter space where $m_\chi \approx M_N$, whereas taking into account the sub-dominant processes does not have a significant impact since they are suppressed by $\frac{M_W^2}{M_N^2}$ and only accessible after electroweak symmetry breaking. For these reasons, we solve the Boltzmann equations numerically for various coupling structures in section \ref{ch:Numerical}. \\
Additionally, we found an analytic solution for the DM relic density in the limit $M_N \gg m_\chi$ where the SM particle scattering processes are in the resonant regime:
\begin{align}
Y_{DM} \left( z \rightarrow \infty \right) = \frac{27}{4 \pi^5 g_\text{eff}^s \sqrt{g_\text{eff}}} \frac{\left( y_\nu y_\chi \right)^2}{y_\nu^2 + y_\chi^2} \frac{M_\text{pl}}{M_N} \, .  \label{eq:DMres}
\end{align}
In case of $y_\chi \ll y_\nu$ we find the observed DM energy density if $y_\chi \approx 10^{-12} \sqrt{\frac{M_N}{m_\chi}}$. \\
However, if $y_\chi \ll y_\nu$ does not hold the approximation of $n_\chi \ll n_\chi^\text{eq}$ we used to derive \eqref{eq:result1} does not apply anymore. To illustrate that we look at the case $y_\chi = y_\nu$, where \eqref{eq:result1} results in:
\begin{align}
Y_{DM} \left( z \rightarrow \infty \right) \approx \frac{3^3 }{2^2 \pi^5 g_\text{eff}^s \sqrt{g_\text{eff}} } \frac{m_\nu M_\text{Pl}}{v^2} \approx 10^{-1} \, .
\end{align}
Using eq. \eqref{eq:eqdens} we find that $Y_{DM}^\text{eq} \lesssim 10^{-2}$. Therefore, $n_\chi \ll n_\chi^\text{eq}$ cannot be satisfied. Hence, the freeze-in scenario does not apply here. Nevertheless, it is still possible to account for the correct amount of DM. In this case, we recover a freeze-out like scenario since due to the resonance the interaction rate becomes as large as the Hubble parameter although the system is feebly coupled. Thus, DM comes into equilibrium with the SM and freezes out as soon as the interaction rate becomes smaller than the Hubble parameter. This occurs approximately at $T=M_N$.\footnote{This is due to the fact that the main contribution to the interaction rate comes from the resonance at $s=M_N^2$, i.e. as soon as the temperature drops below $M_N$ the resonance cannot be reached efficiently anymore and therefore the interaction rate decreases significantly.}   
%However, for the cases of $y_\nu \approx y_\chi$ and $y_\nu \ll y_\chi$ the bracket in eq. \eqref{eq:DMres} approximately results in
%\begin{align}
%2^7 \pi y_\nu^2 + y_\chi^4 \, .
%\end{align}
%We have $y_\chi^4 \gtrsim 2^7 \pi y_\nu^2$ as long as $m_\chi \lesssim 10 \, \mathrm{eV}$\footnote{Otherwise DM would be overproduced since the contribution of $y\chi^4$ would be too large}. If $m_\chi \lesssim 10 \, \mathrm{eV}$ a relic density of $Y_\text{DM} \gtrsim 10^{-2}$ is required to fit the observed DM energy density. This, however, contradicts the assumption of $n_\chi \ll n_\chi^\text{eq}$. Thus, for this region eq. \eqref{eq:DMres} is not the correct result. \\ Moreover, if $2^7 \pi y_\nu^2 \gtrsim y_\chi^4$ which applies for $y_\nu \approx y_\chi$ we find that the produced relic density does not depend on the heavy neutrino mass and that the DM mass is required to be of $m_\chi \approx \mathcal{O} \left( eV \right)$. Therefore, also in this case our approximation does not apply and the result is not correct. \\ 
 Consequently, the number density can be estimated by the equilibrium density at freeze-out: 
 %and since we assumed $M_N \gg m_\chi$ the number density \eqref{eq:eqdens} at freeze-out yields:
\begin{align}
Y_\text{DM} \left( z \rightarrow \infty \right) = Y_\chi^\text{eq} \left( T \approx M_N \right) \overset{M_N \gg m_\chi}{=} \frac{45 g_\chi }{2 \pi^4 g_\text{eff}^s} \approx 10^{-3} \, . 
\end{align} 
Equating this result with eq. \eqref{eq:DMexp} yields a DM mass of $m_\chi = \mathcal{O} \left( 100 \, \mathrm{eV} \right)$. In contrast to the non-resonant case, this DM mass violates the Tremaine–Gunn bound which restricts fermionic DM to have a mass of at least roughly a $\mathrm{keV}$ \cite{Tremaine:1979we}. Therefore, DM must be bosonic in this case. However, this case also is in tension with observations of the Lyman-$\alpha$ forest which allows to probe structures of the size $10^{0-2} h^{-1} \, \mathrm{Mpc}$ \cite{Baur:2017stq}. This issue is treated in more detail within chapter \ref{ch:Bounds}.         \\
We summarized our results for the case $y_\chi = y_\nu$ in a schematic plot (see fig. \ref{fig:niceplot}).
%Finally, we would like to comment on further possible implications of the model, namely higgs decays and direct detection. \\
%In case of $M_N<m_h$, the decay $h \rightarrow \nu N$ is allowed on tree level, resulting in:
%\begin{align}
%\Gamma_{h \rightarrow \nu N} = \frac{y_\nu^2}{8 \pi} m_h \left( 1- \frac{M_N^2}{m_h^2} \right)^2 = \frac{m_\nu M_N m_h}{8 \pi v^2} \left( 1- \frac{M_N^2}{m_h^2} \right)^2 \lesssim 10^{-13} m_h \, .
%\end{align}
%In case of $M_N > m_h$, we expect an even smaller contribution since only a three body decay, e.g. $h \rightarrow Z \nu \nu$ would be allowed. Those are proportional to $y_\nu^4 \frac{M_W}{M_N}$ instead of $y_\nu^2$. Thus, the additional heavy neutrino does not affect the higgs decay width significantly. \\
%Concerning direct detection we refer to
%As before, considering a model which generates $y_\chi \approx y_\nu = \frac{\sqrt{m_\nu M_N}}{v}$ increases the predictivity of the model and results in
%\begin{align}
%2^7 \pi y_\nu^2 + y_\nu^4 = 2^7 \pi \frac{m_\nu M_N}{v^2} + \left( \frac{m_\nu M_N}{v^2} \right)^2 \approx 10^{-21} \frac{M_N}{m_\chi}
%\end{align}
%As long as $M_N \lesssim 10^{15} \, \mathrm{GeV}$ the second term is negligible and the produced relic denisity does not depend on the heavy neutrino mass and therefore the DM mass is required to be of $m_\chi \approx \mathcal{O} \left( eV \right)$.  

\section{Numerical Analysis} \label{ch:Numerical}
\begin{figure}
 	\begin{center}
 		\includegraphics[width=105mm]{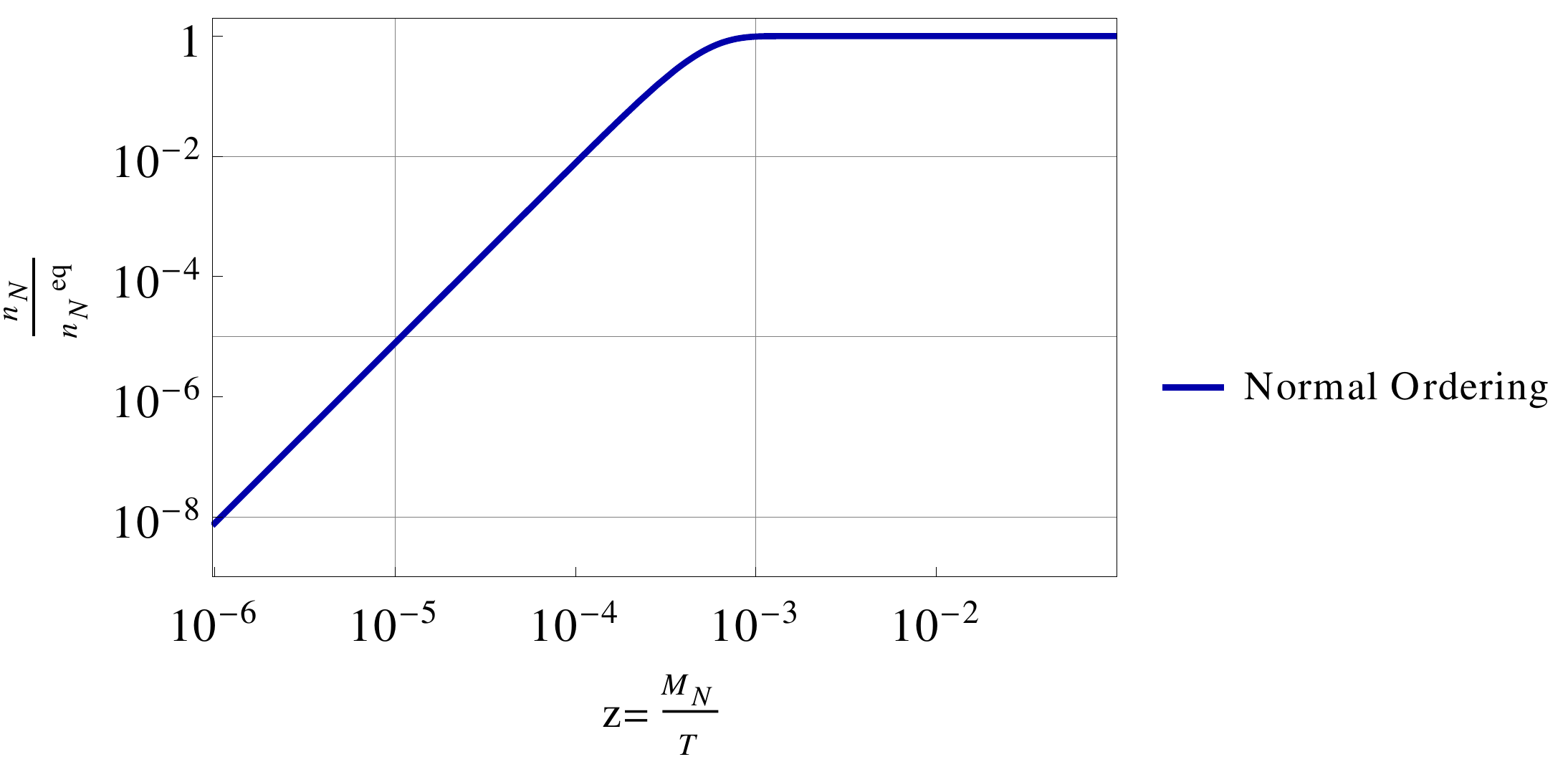}
 	\end{center}
 	\caption{The ratio of the heavy neutrino density to its equilibrium density against the dimensionless quantity $z= M_N T^{-1}$ in case of normal ordering. The heavy neutrinos reach equilibrium for $T \approx 10^3 M_N$.  }
 	\label{fig:Neq}
 \end{figure}

We solved the Boltzmann equations numerically in the non-resonant case for different coupling structures $y_\chi = (0.1,1,10) y_\nu$ and DM masses of $m_\chi \in [10^2 , 10^{10}]\, \mathrm{GeV}$ assuming different flavor structures, i.e. $f_1 \left( \theta \right) = (10^{-1},1)$ and $f_2 \left( \theta \right) = 2.46$. \\
Since we investigate a feebly coupled sector, the back reactions in the DM production processes can be neglected. Only for the processes $N \leftrightarrow \nu h$ responsible for producing the mediator $N$ the back reactions are relevant, since for most of the parameter space $N$ equilibrates with the SM. \\
Therefore, we solve the Boltzmann equation in two steps:
\begin{enumerate}
	\item
	 The $N$ production via $N \leftrightarrow \nu h$ is solved at the level of the momentum distribution function, thereby taking into account the non-thermal shape of the distribution. The details of solving the Boltzmann equations at the level of momentum distribution functions are given in appendix \ref{app:boltzmann} and the collision term for the process in question is given in eq. \eqref{eq:ColN}. Eventually, this procedure results in the quantity $\frac{n_N}{n_N^\text{eq}} \left( T \right)$. We take $\frac{n_N}{n_N^\text{eq}} \left( T \rightarrow \infty \right) = 0$ as initial condition. 
	 \item 
	 This quantity is used to solve the Boltzmann equations for DM production via heavy neutrino and SM particle scattering employing the formalism described in chapter \ref{ch:Boltzmann}. We take vanishing number densities for the DM particles as initial conditions. The SM particles are assumed to follow their equilibrium densities throughout the production process. The final result is then given by $Y_\text{DM} = Y_\chi + Y_\phi$ for $T \rightarrow 0$. Note that the independent solution of the Boltzmann equations for the dark sector particles and the heavy neutrino is only possible due to the tiny interaction rate, which allows to neglect the back reactions from DM production via heavy neutrino scattering.   
	 \end{enumerate} 
 \begin{figure}
 	\begin{center}
 		\includegraphics[width=105mm]{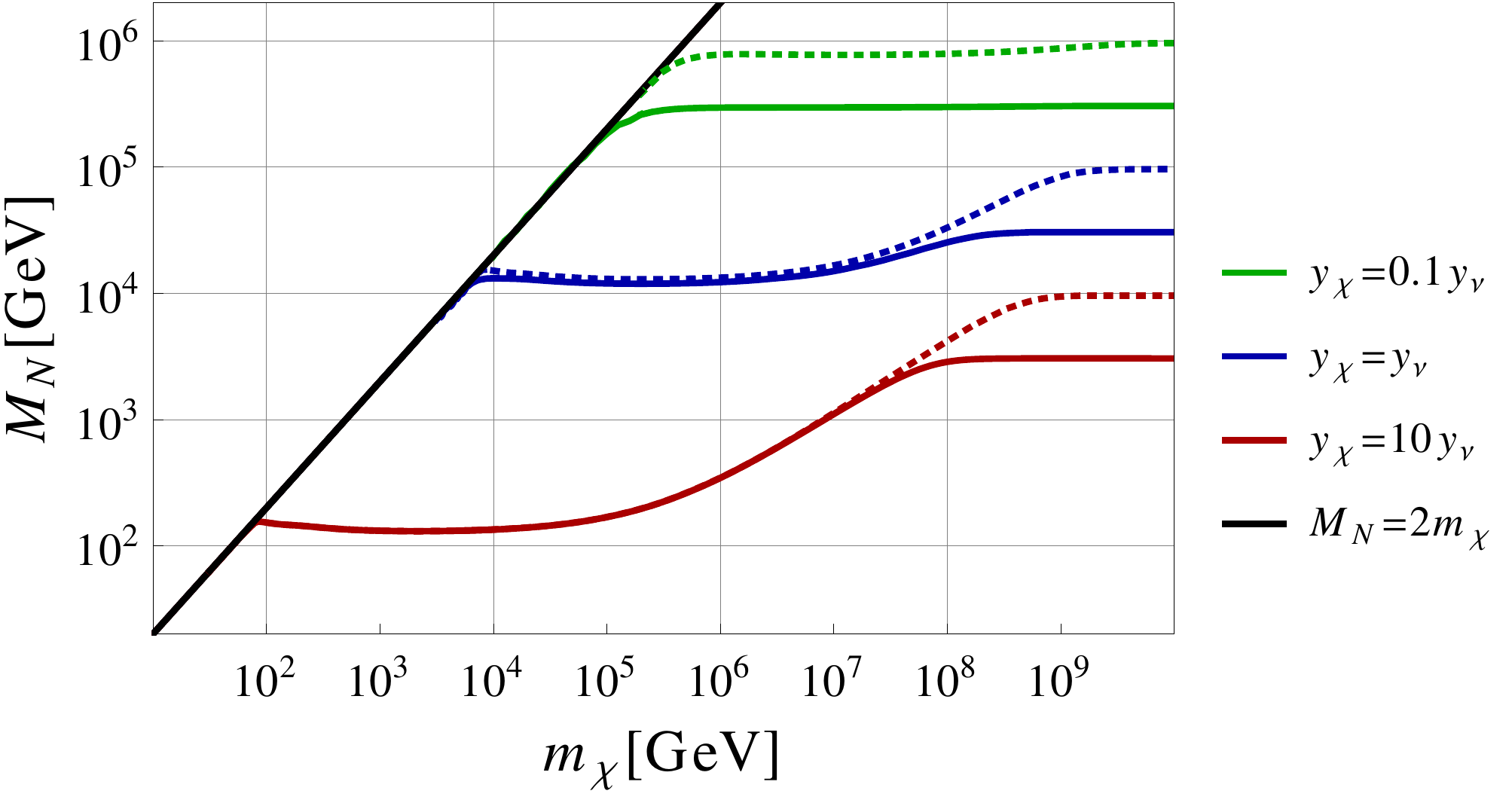}
 	\end{center}
 	\caption{The numerically obtained DM density $Y_\text{th}$ is compared to the observed DM density $Y_\text{exp}$ for different values of the DM mass $m_\chi$ and the mediator mass $M_N$: The different colored solid lines represent the points where the observed DM density is reproduced for a certain coupling structure. A parameter point above a specific line overproduces DM for the corresponding coupling structure while points below do not generate enough DM. Lines of the same color have the same coupling structure. A solid line represents a scenario with of $f_1 \left( \theta \right) = 1$, while a dotted represents a scenario with $f_1 \left( \theta \right) = 0.1$. The black line separates the plane into the non-resonant (lower right) and resonant (upper left) regime. The latter one was not scanned.  }
 	\label{fig:DMDplot}
 \end{figure}
The results are summarized within figure \ref{fig:DMDplot}. From our earlier considerations in chapter \ref{ch:DisAna} we expect the setup to work for a constant mediator mass $M_N$ as long as $m_\chi \gg M_N$. This constant value can be obtained by solving eq. \eqref{eq:result2} for a given coupling structure. Consider e.g. the case $y_\nu= y_\chi$, where eq. \eqref{eq:result2} results in $M_N \approx 10 \, \mathrm{TeV}$. This case is illustrated by the solid blue line in figure \ref{fig:DMDplot}. For $10 \, \mathrm{TeV} \leq m_\chi \leq 10^4 \, \mathrm{TeV}$ the prediction is met by the numerical solution. For larger DM masses, however, a larger mediator mass is required to accommodate the observed relic density. This is due to the following reason: The freeze-in mechanism produces DM efficiently down to temperatures around the heaviest mass involved in the production process. For the non-resonant regime this mass is given by the DM mass itself. Therefore, DM production is efficient for $T \gtrsim m_\chi$. The mediator mass and, thereby the neutrino Yukawa $y_\nu$, start to increase as soon as $\frac{n_N}{n_N^\text{eq}} \left( T \right) \ll 1$ for $T \gtrsim m_\chi$, since this suppresses DM production via heavy neutrino scattering. In case of $y_\nu = y_\chi$ heavy neutrino scattering accounts for $\frac{35}{41}$ of the DM production if the heavy neutrinos are following their equilibrium density during the time of production. If this contribution is missing, it has to be compensated by a larger neutrino Yukawa which results in a larger mediator mass. \\
The heavy neutrinos reach thermal equilibrium with the SM for $T \sim c M_N$. The factor $c$ is independent of the neutrino Yukawa $y_\nu$ and in case of normal ordering is independent of the parameters $\theta$ which encode the flavor structure of the neutrino Yukawas. The evolution of the heavy neutrino number density is shown in figure \ref{fig:Neq}. Here, the heavy neutrinos reach equilibrium for $T \approx 10^{-3} M_N$. Therefore, the lines in figure \ref{fig:DMDplot} start to deviate significantly from a constant value of $M_N$ for $m_\chi > 10^3 M_N$, since in this case it is $\frac{n_N}{n_N^\text{eq}} \left( T \right) < 1$ for the complete production time. A constant value of $M_N$ is reached again if the contribution of the heavy neutrino scattering becomes negligible. \\
For $f_1 = 0.1$ the contribution of SM particle scattering is suppressed by a factor of 10 since the contribution of the SM particle scattering is proportional to $f_1$. Thus, a larger coupling compared to $f_1 = 1$ is required. This effect can be seen in figure \ref{fig:DMDplot} where all dotted lines lie above the solid line of the same color. \\
The different couplings structures result in larger (smaller) mediator masses for a small (large) dark Yukawa coupling compared to the neutrino Yukawa. Additionally, the effect of a small $f_1$ differs for a small (large) dark Yukawa. While the increase with a larger DM mass becomes less significant for a small dark Yukawa, the absolute difference between the small and large $f_1$ cases becomes stronger. This is due to the different contributions from heavy neutrino and SM particle scattering for the different coupling structures. \\
For smaller DM masses close to the transition to the resonant regime, the correct DM relic density is obtained for values of $M_N$ very close to $M_N = 2 m_\chi$. Although not visible within figure \ref{fig:DMDplot}, all lines follow the black line down to small DM masses until the enhancement close to the resonance is not strong enough anymore to generate a sufficient amount of DM. However, the numerical solution is not trustworthy in this area due to numerical instabilities and therefore not presented here. We estimate the lower bound on $m_\chi$ by evaluating eq. \eqref{eq:resultnr} in the limit $M_N \rightarrow 2 m_\chi$. In the case of $y_\chi = \alpha y_\nu$ we obtain $m_\chi \gtrsim \alpha^{- \frac{4}{3}} \, \mathrm{MeV}$. 

\section{Constraints} \label{ch:Bounds}
In this section we discuss different constraints on the model. At first we discuss constraints from structure formation which pose strong limits in the resonant regime. Afterwards we investigate the impact of direct detection bounds on our parameter space and briefly discuss charged lepton flavor violation and indirect detection.
\subsection{Structure Formation} 
Since DM particles only interact weakly with the SM they can escape from gravitational wells formed in the early universe, thereby delaying structure formation below their free-streaming scale. Given the redshift at the production time $z_\text{prod}$ the free-streaming scale is given by
\begin{align}
\lambda_{fs} = \int \limits_0^{z_\text{prod}} dz \frac{v \left( z \right)}{H \left( z \right)  } \label{eq:freestreaming} \, ,
\end{align}
where $v \left( z \right)$ is the DM velocity at a given redshift $z$. \\
The observation of absorption lines in the spectra of distant quasars mostly induced by hydrogen clouds, the so called Lyman-$\alpha$ forest, allows for probing structures on the scale of roughly $10^{0-2} h^{-1} \mathrm{Mpc}$\cite{Baur:2017stq}. \\
Following the lines of \cite{Garny:2018ali}, we estimate the free-streaming scale for the case of DM in equilibrium with the SM up to a certain freeze-out temperature and for the case of resonantly produced DM still in the freeze-in regime. Within this model, the first case applies to the resonant production with a coupling structure of $y_\nu \lesssim y_\chi$ whereas the latter is present in the resonant production regime for $y_\chi \ll y_\nu$. The non-resonant production regime is not investigated here due to the much larger DM masses that are required to generate the observed relic density. Therefore, we do not expect this case to be in tension with the Lyman-$\alpha$ forest. \\
As it was pointed out in \cite{Konig:2016dzg}, the free-streaming scale should only be understood as an order-of-magnitude estimator in the case of non-thermal DM momentum distribution and may differ up to $\mathcal{O} \left( 1 \right)$ factors from results obtained with dedicated tools like the CLASS-code which computes the  linear matter power spectrum. \\
 For the purposes of this work, the estimation of the free-streaming length suffices, firstly because the non-thermal momentum distribution produced by the resonant freeze-in process (eq. \eqref{eq:mdf}) is close to a thermal shape and secondly because the resonantly produced DM for the freeze-out case will be excluded by this method by roughly two orders-of-magnitude. \\
We approximate the velocity in eq. \eqref{eq:freestreaming} by the average velocity at the production time $z_\text{prod}$ which is only redshifted afterwards, i.e. 
\begin{align}
v \left( z \right)  =  \frac{p \left( z \right)}{\sqrt{p \left( z\right)^2 + m_\chi^2}} \, ,  
\end{align}
with 
\begin{align}
p \left( z\right) = p_\text{prod} \frac{1 + z}{1 + z_\text{prod}} \, ,
\end{align}
and 
\begin{align}
p_\text{prod} = \frac{\int dp \, p^3 f \left( p, T_\text{prod} \right)}{\int dp \, p^2 f \left( p,T_\text{prod} \right)} \, .
\end{align}
Moreover, the Hubble Parameter is given by 
\begin{align}
H \left( z \right) = H_0 \sqrt{ \Omega_m \left( 1 + z \right)^3 + \Omega_r \left( 1 + z \right)^4 + \Omega_\Lambda} \, .
\end{align}
For the numerical evaluation, we use the cosmological parameters of \cite{Aghanim:2018eyx}. Lastly, we use the relation between the temperature and the redshift $T = T_0 \left(1 + z\right) \left( \frac{g_s^\text{eff} \left( T_0 \right)}{g_s^\text{eff} \left( T \right)}\right)^\frac{1}{3}$ to give $T_\text{prod}$ in terms of the redshift. The temperature $T_0$ refers to the temperature today. Inserting these expressions into \eqref{eq:freestreaming} allows for calculating $\lambda_{fs}$ in terms of the production time $z_\text{prod}$ and the average momentum at this time $p_\text{prod}$. Then, the result is compared to the upper bound on the free-streaming scale of $\lambda_{fs} \lesssim 0.1 \, \mathrm{Mpc}$ which was derived in \cite{Garny:2018ali} assuming that the particle species in question accounts for all of the observed DM relic density. \\
In case of resonant production with $y_\nu \lesssim y_\chi$ we can assume DM to have a Boltzmann like momentum distribution, i.e. $f\left(p,T \right) = \exp \left( - E_p T^{-1} \right)$. We take the time of production to be the freeze-out temperature since the interactions of DM with the SM cease to be efficient from this point on. For this distribution the average momentum results in
\begin{align}
p_\text{prod} = \frac{m_\chi^2 + 3 m_\chi T_\text{prod} + 3 T_\text{prod}^2}{m_\chi + T_\text{prod}} \, .
\end{align} 
By comparing the interaction rate $\Gamma$ of the process $vh \rightarrow \chi \phi$ in the resonant regime to the Hubble parameter we find that $T_\text{prod} \approx M_N$. For mediator masses $M_N \gtrsim \mathrm{MeV}$ the free-streaming scale becomes insensitive to the mediator mass itself beside the change induced by the different $g_S^\text{eff} \left(T_\text{prod} \right)$. In this case, we find a lower bound on the DM mass of $m_\chi \gtrsim 10 \, \mathrm{keV}$. However, we found in chapter \ref{ch:DisAna} that a DM mass of $0.1 \, \mathrm{keV}$ is required in order not to overproduce DM within this scenario. This lies two orders of magnitude below the estimated lower bound. Therefore, the resonant production regime with $y_\nu \lesssim y_\chi$ is excluded by the Lyman-$\alpha$ measurement. \\
If, on the other hand, $y_\chi \ll y_\nu$, DM does not equilibrate with the SM even in the resonant production regime. Therefore the spectrum is non-thermal and given by eq. \eqref{eq:mdf}. We take $z_\text{prod} \left( T_\text{prod} \right)$ as the temperature where the derivative of the total particle number with respect to the time is maximized. Therewith, we find $T_\text{prod} = 3.36 M_N$ which results in $p_\text{prod} = 0.4 T_\text{prod}$. Here, we also find that for $M_N \gg m_\chi$ the free-streaming scale is insensitive to the mediator mass and the lower bound on the mass results in $m_\chi \gtrsim 3 \, \mathrm{keV}$. \\
To summarize, the Lyman-$\alpha$ measurement strongly constraints the resonant production regime of this model. While the case where the resonant enhancement of the production cross section is strong enough to equilibrate DM with the SM is completely ruled out, the freeze-in regime is only allowed for couplings $y_\chi \lesssim 10^{-12} \sqrt{\frac{M_N}{\mathrm{keV}}}$ with $m_\chi \gtrsim 3 \, \mathrm{keV}$.
\subsection{Direct Detection}
\begin{figure}
	\centering
		\includegraphics[width=70mm]{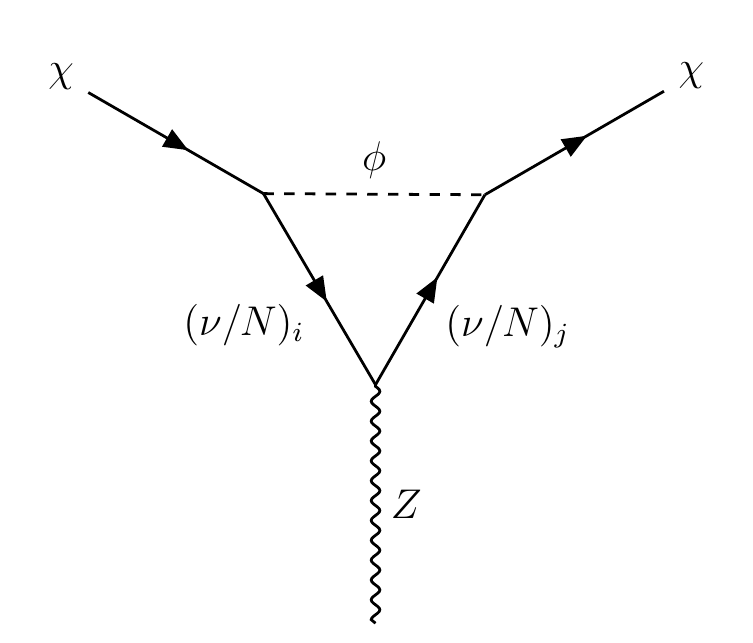}
%	\feynmandiagram[medium,vertical=e to f] {a [particle=\(\chi\)] -- [fermion] b -- [scalar,edge label=\(\phi\)] c -- [fermion] d [particle=\(\chi\)],b -- [fermion,edge label'=\((\nu/N)_i\)] e -- [fermion,edge label'=\((\nu/N)_j\)] c,e -- [boson,edge label=\(Z\)]  f;
%	}; 
\caption{1-Loop diagram generating the effective coupling of DM to the Z. The indices $i,j$ run from 1 to 3.}\label{fig:Zloop}
\end{figure}

\begin{figure}
	\centering
	\includegraphics[width=115mm]{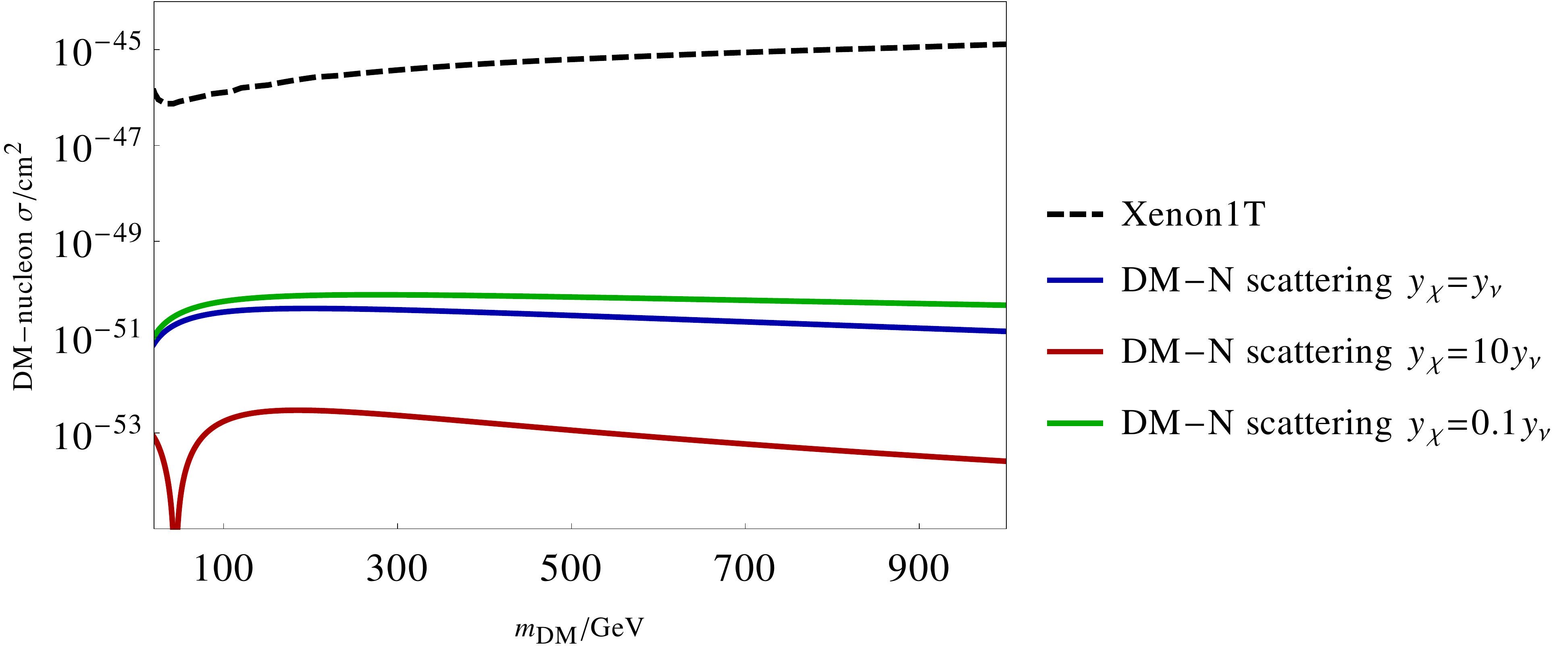}
	\caption{The expected direct detection signals for the coupling structures investigated within chapter \ref{ch:Numerical} are compared to the current bounds from XENON1T \cite{Aprile:2017iyp} (dashed black curve). The dip in the red curve is due to a cancellation appearing in the loop function.} 
	\label{fig:XENON}
\end{figure}

Direct detection experiments search for interactions of DM with nuclei. In this model, a coupling of DM to the $Z$ boson is generated at one loop. The corresponding Feynman diagram is shown in figure \ref{fig:Zloop}. 
The coupling to the $Z$ is then given by $\mathcal{L} \supset g_{Z \chi \chi} \bar{\chi} \gamma^\mu P_L \chi Z_\mu $ with \cite{Chowdhury:2018nhd}
%\begin{align}
%g_{Z \chi \chi} = &- \frac{y_\chi^2}{16 \pi^2} \frac{g_w}{4 \cos \theta_w} \frac{v^2}{M_N^2} \sum \limits_{k,m = 1}^3 \left( Y_\nu^T Y_\nu \right)_{km} \left[ M_N^2 C_0 \left( m_\phi , M_N , M_N \right) + \right.  \nonumber \\ &\left. 2 \left( C_{24} \left( m_\phi , M_N , M_N \right) + C_{24} \left( m_\phi ,0 , 0 \right) -C_{24} \left( m_\phi , M_N , 0 \right) -C_{24} \left( m_\phi , 0 , M_N \right) \right)  \right] \, . 
%\end{align}
%The expression for this coupling is taken from eq. (25) of \cite{Chowdhury:2018nhd}. The loop functions $C_0$ and $C_{24}$ are defined as in the appendix of \cite{Chowdhury:2018nhd}. For our model we have $g_{Z \phi} = g_l^R =0$ in eq.(25) but have six different fermions propagating in the loop, namely the six neutrino mass eigenstates. By simplifying the sum of the loop functions we obtain
\begin{align}
g_{Z \chi \chi} = &- \frac{y_\chi^2}{16 \pi^2} \frac{g_w}{4 \cos \theta_w} \frac{\Delta m_\nu}{M_N}  2.3 \cdot g \left( \frac{M_N^2}{m_\phi^2} \right) \, ,
\end{align}
and
\begin{align}
g \left( x \right) = \frac{x \left[\left(x + 2\right) \log \left( x \right) + 3 \left(1 - x\right)\right]}{2 \left(1 -x \right)^2} \, ,
\end{align}
where we have used the best fit values of \cite{Esteban:2018azc} for the parameters of the PMNS matrix in case of normal ordering, which yields $\sum \limits_{k,m = 1}^3 \left( Y_\nu^T Y_\nu \right)_{km} \approx 2.3 \cdot y_\nu^2$. \\
Therewith, DM interacts with quarks via $Z$ exchange. Since this process happens at energies far below the $Z$ mass, the heavy mediator is integrated out leading to 
\begin{align}
\mathcal{L} \supset \frac{1}{M_Z^2} \left[ g_{Z \chi \chi} \bar{\chi} \gamma^\mu \left( 1 - \gamma^5 \right) \chi \right] \left[ \bar{q} \gamma_\mu \left( g_{q v} + g_{q a} \gamma^5 \right) q \right] \, ,
\end{align}
where $g_{q v}$ and $g_{q a}$ are the couplings of the SM quarks to the $Z$. At low energies only the vector-vector and axial-axial interactions are not suppressed by powers of the relative velocity or momentum transfer, thereby leading to  a spin-independent and a spin-dependent DM-nuclei cross section, respectively \cite{Escudero:2016gzx,Berlin:2014tja}. For the spin-independent cross section we obtain \cite{Berlin:2014tja} 
\begin{align}
\sigma_{SI} = \frac{\mu_{\chi N} g_{Z \chi \chi}^2}{\pi M_Z^4} \left[ Z \left( 2 g_{uv} + g_{dv}\right) + \left(A - Z\right) \left( g_{uv} + 2 g_{dv} \right) \right] \, ,
\end{align}
with $\mu_{\chi N} = \frac{m_\chi M_{Xe}}{m_\chi + M_{Xe}}$, $g_{uv} = g_w \left( \frac{1}{4 \cos \theta_w} - \frac{2 \sin^2 \theta_w}{3 \cos \theta_w}\right)$ and $g_{dv} = g_w \left( - \frac{1}{4 \cos \theta_w} + \frac{ \sin^2 \theta_w}{3 \cos \theta_w}\right)$. \\
This cross section is constrained by the XENON experiment, as shown in figure \ref{fig:XENON}. Therefore, the freeze-in setup cannot be constrained by this measurement. There are scenarios considered in the literature which allow for having a large direct detection signature even in a freeze-in scenario \cite{Hambye:2018dpi}. In \cite{Hambye:2018dpi}, the cross section is enhanced due to a very light mediator. Since in our model the interaction is mediated by a $Z$ boson this does not apply here. 
\subsection{Indirect Detection and HEP Phenomenology}  
Prospects for indirect detection of DM such as the observations of $\gamma$-rays from the galactic center or the precise measurement of the CMB all rely on the efficient annihilation of DM into SM particles. In the case of neutrino portal DM this usually happens subsequently by DM first annihilating into heavy neutrinos which then decay or annihilate into SM particles. Several prospects for indirect detection were investigated in \cite{Batell:2017rol} for the case of freeze-out production of DM where before DM freezes out its annihilation is efficient. This, however, is not the case for the freeze-in scenario investigated in this work. Here, the process is efficient only in the direction of DM production. This leads to a suppression of the annihilation cross section $<\sigma v>$ which enters all observables of indirect detection considered in \cite{Batell:2017rol} since the couplings $y_{\nu}$ and $y_\chi$ are required to be feeble. Moreover, the annihilation rate is suppressed by a factor $\frac{n_\text{DM}}{n_\text{DM}^\text{eq}}$ compared to the freeze-out case. For this reason we do not study indirect detection observables within this work. \\
The minimal version of the type I seesaw mechanism employed here induces couplings of the SM gauge bosons and the Higgs to the heavy neutrino states. This can modify electroweak precision observables and induce charged lepton flavor violation (LFV) as well as additional Higgs decay channels in case of a light heavy neutrino\cite{Antusch:2014woa,Molinaro:2013toa}. The strongest constraints come from the decay $\mu \rightarrow e \gamma$ with $\mathcal{B} \left( \mu \rightarrow e \gamma \right) \leq 4.2 \cdot 10^{-13}$ \cite{Tanabashi:2018oca}. Within this setup the decay is mediated at one loop level by a $W$ boson and a neutrino. The branching ratio of this process is then given by \cite{Antusch:2006vwa}:
\begin{align}
\frac{\Gamma \left( \mu \rightarrow e \gamma \right)}{\Gamma \left( \mu \rightarrow \nu_\mu e \bar{\nu}_e \right)} = \frac{3 \alpha}{32 \pi} \frac{|\sum \limits_{k=1}^6 U_{\mu k} U^\dagger_{e k} F \left( x_k \right)|^2 }{\sum \limits_{k,j=1}^3 U_{\mu k} U^\dagger_{\mu k} U_{e l} U^\dagger_{e l}  } \, , 
\end{align}
where $F \left( x_k \right)$ is a loop function with $x_k = m_k^2 M_W^{-2}$. Since we assumed the heavy neutrinos to be mass-degenerate and the light neutrino mass is tiny compared to $m_W$ we split the sum in the numerator into two parts with $F \left( 0 \right) = \frac{10}{3}$ and $F \left( \frac{M_N^2}{M_W^2} \right)$. Additionally we neglect the small deviation from one in the diagonal elements of $U_\text{PMNS} U_{PMNS}^\dagger$ in the denominator. Since the mixing matrix $U$ is unitary we find
\begin{align}
\frac{\Gamma \left( \mu \rightarrow e \gamma \right)}{\Gamma \left( \mu \rightarrow \nu_\mu e \bar{\nu}_e \right)} = \frac{3 \alpha}{32 \pi} \frac{\Delta m_\nu^2}{M_N^2} \left( F \left( 0 \right) -F \left( \frac{M_N^2}{M_W^2} \right) \right)^2 |\left( U_\text{PMNS} \frac{m_\nu}{\Delta m_\nu} U_\text{PMNS}^\dagger \right)_{\mu e}|^2 \, .    
\end{align} 
Taking the best fit values from \cite{Esteban:2018azc} we find $ \left( U_\text{PMNS} \frac{m_\nu}{\Delta m_\nu} U_\text{PMNS}^\dagger \right)_{\mu e} = 0.12$. Thus, we can give the branching ratio as a function of the heavy neutrino mass only since the free parameters of the orthogonal matrix $R$ cancel within this setup \cite{Casas:2001sr}. This expression is maximized for $M_N = 1.36 M_W$ and results in
\begin{align}
\frac{\Gamma \left( \mu \rightarrow e \gamma \right)}{\Gamma \left( \mu \rightarrow \nu_\mu e \bar{\nu}_e \right)} =  \frac{3 \alpha}{32 \pi} \frac{\Delta m_\nu^2}{M_W^2} 0.12^2	\cdot 0.266 \approx 10^{-31} \, , 
\end{align}
which is far below the experimental limit. For this reason, we also expect other LFV and electroweak precision observables not to significantly constrain the scenario. \\
Another imprint of this model could be found in additional decay channels of the Higgs if $M_N < m_h$. In this case the decays $h \rightarrow \nu_i N_j $ and $h \rightarrow N_i N_j$ are kinematically accessible. As pointed out in \cite{Escudero:2016ksa,Gago:2015vma} the dominant contribution comes from the decay into a heavy and a light neutrino. However, branching ratios of this process larger than $10^{-2}$ are already ruled out and are typically much smaller due to the tiny Yukawa coupling\cite{Gago:2015vma}. Therefore, the contribution is negligible. 
\section{Conclusion} \label{ch:conlusion}
We have investigated a minimal neutrino portal DM model. The SM is extended by three right-handed neutrinos which generate the neutrino masses via a type I seesaw mechanism and, furthermore, act as mediator between the SM and DM. The dark sector consists of a boson $\phi$ and fermion $\chi$ coupled to the right handed neutrino via a Yukawa coupling. Motivated by the small Yukawa couplings of the type I seesaw mechanism in case of small heavy neutrino masses of $M_N \lesssim \mathcal{O} \left( \mathrm{PeV} \right)$ we studied DM production via the freeze-in mechanism. \\
We derived analytic solutions for the number density in the resonant ($M_N > m_\chi + m_\phi$) and non-resonant ($M_N < m_\chi + m_\phi$) DM production regime. Adding the requirement that the coupling of the right-handed neutrino to the SM is of the same order of magnitude as its coupling to the dark sector allows for the prediction of the mediator or the DM mass respectively. In the non-resonant regime, we find $M_N \approx 10 \, \mathrm{TeV}$. The non-resonant regime is studied in more detail numerically, as seen in figure \ref{fig:DMDplot}.  \\
Within the resonant regime, however, for $y_\chi \gtrsim y_\nu$ the resonant production of DM is strong enough to bring it into equilibrium with the SM. Thus, the freeze-out mechanism is revovered although the couplings between DM and the SM are feeble. Moreover, in this scenario we can predict a DM mass of $m_\chi \approx 100 \, \mathrm{eV}$. For $y_\chi \ll y_\nu$, nonetheless, DM production via freeze-in is still possible. To satisfy the observed DM energy density the coupling of the right-handed neutrino to DM is required to be $y_\chi \approx 10^{-12} \sqrt{\frac{M_N}{m_\chi}}$. \\
The resonant scenario is strongly constrained by the measurement of the Lyman-$\alpha$ forest. The freeze-out case can be excluded completely, while freeze-in with $y_\chi \approx 10^{-12} \sqrt{\frac{M_N}{m_\chi}}$ is only viable for $m_\chi \gtrsim 3 \, \mathrm{keV}$. Charged lepton flavor violation, Higgs decays, indirect detection and direct detection have little impact on our parameter space due to the feeble coupling of the SM to the dark sector. 
Thus, producing the observed DM energy density within this model of neutrino portal DM is possible even with small couplings between the SM and the dark sector. \\
Although within this work CP violation in the PMNS matrix was assumed to be absent, it could be included in the analysis to explore its phenomenological imprints and its impact on leptogenesis. 
%Moreover, the vacuum stability of the scalar potential and the stability of DM might be spoiled by the running of the scalar mass $m_\phi^2$ and the four vertex coupling $\lambda \phi^4$ induced by the coupling of the scalar to the fermions $\nu_R$ and $\chi$. Since these effects are typically stronger for heavy particles in the loop this may offer a possibility to constrain the non-resonant production regime. However, this requires a more dedicated analysis and lies beyond the scope of this paper and will be subject of a future work. 
%An interesting extension of this model would be the explanation of the small Yukawa couplings of the right-handed neutrino to the SM and DM by a suppression mechanism. One explanation could be an extra dimensional model where the heavy neutrino in contrast to the SM and DM particles propagates in an extra dimension since it is the only singlet under all considered gauge groups. Thus, all couplings to the right-handed neutrino are suppressed by a volume factor which can even generate $y_\chi \approx y_\nu$. \\
%Due to the tiny couplings of the right-handed neutrino to SM and DM direct detection seems not viable. Therefore, exploring the detectability of this model might be an interesting prospect for a future work. Moreover, we only demonstrated a few working cases of this model. For example we considered only the case of $m_\chi = m_\phi$ and $M_N > M_W$. Thus, an exploration of $M_N < M_W$ which leads to different dominant interactions between the heavy neutrino $N$ and the SM or a hierarchic dark sector might be interesting.   
    
\section*{Acknowledgments}
I would like to thank Prof. H. Päs for providing the possibility to work on this project and his constant support throughout the work. 

\appendix
\section{Boltzmann Equations at the Level of Momentum Distribution Functions}
\label{app:boltzmann}
A common simplifying assumption (e.g. in \cite{Giudice:2003jh}) to solve the Boltzmann equation is to perform the momentum integration by assuming that if a particle distribution deviates from its equilibrium density it differs only by a momentum-independent factor, i.e. $f_i = \alpha_i f_i^\text{th}$ with $\frac{\partial \alpha_i}{\partial p_i} = 0$. Furthermore, the equilibrium densities of bosons and fermions are approximated by a Boltzmann distribution. \\
Following the lines of \cite{Konig:2016dzg,Heeck:2017xbu} we solve the Boltzmann equations at the level of momentum distribution functions. This has the advantage of a more accurate solution and the exact shape of the momentum distribution allows for more insights into the process of structure formation. Throughout the calculation we approximate the equilibrium densities of any particle species by a Boltzmann distribution. The Boltzmann equation is given by:
\begin{align}
\left( \frac{\partial}{\partial t} - H p \frac{\partial}{\partial p} \right) f \left( p , T \left( t \right) \right) = \mathcal{C} \left( p , T \right) \, . \label{eq:boltzmannmom}
\end{align}
Here $t$ is the time, $H$ the Hubble parameter, $f$ is the momentum distribution function of the particle species whose evolution is described by this Boltzmann equation, $p$ is their momentum and $\mathcal{C} \left( p , T \right)$ is the collision term which describes the impact of interactions. For the integration of this equation it is convenient to perform a coordinate transformation $(t , p) \rightarrow (r, x)$ such that the differential operator on the left hand side contains a derivative with respect to one of the new variables only. If $r$ only depends on $t$ and 
\begin{align}
\frac{\partial x}{\partial t} - H p \left( r, x \right) \frac{\partial x}{\partial p} = 0 \, , \label{eq:condition1}
\end{align} 
the L.H.S. of eq. \eqref{eq:boltzmannmom} results in
\begin{align}
\frac{\partial r}{\partial t} \frac{\partial }{\partial r} \, . 
\end{align}
The condition \eqref{eq:condition1} is fulfilled if 
\begin{align}
x \left( p, t \right) = x \left( \frac{a \left( t \right)}{a \left( t_0 \right)} p , t_0 \right)
\end{align}
A convenient choice for $x$ is 
\begin{align}
x \left( p , t \right)  = \frac{1}{T_0}  \frac{a \left( t \right)}{a \left( t_0 \right)} p = \left( \frac{g_s \left( T_0 \right)}{g_s \left( T \right)} \right)^{\frac{1}{3}} \frac{p}{T} \, . \label{eq:x}
\end{align}
For the last equality we used the conservation of entropy $s(T_0) a(T_0) = s(T) a(T) = \text{const.}$ and $g_s$ are the entropy degrees of freedom.  The conservation of entropy also allows us to relate the temperature $T$ to the time $t$: 
\begin{align}
\frac{dT}{dt} = - H T \left( 1 + \frac{T}{3} \frac{d g_s}{dT} g_s^{-1} \right)^{-1} \, .
\end{align} 
Since $T$ is only a function of $t$ and not of $p$ we can choose 
\begin{align}
r \left( T \right) = \frac{m_0}{T} \, , \label{eq:r}
\end{align}
with $m_0$ being am arbitrary mass scale. Combining all this the Boltzmann equation results in 
\begin{align}
r H \left( 1 - \frac{T}{3} \frac{\partial}{\partial r} \ln \left( g_s \right)  \right)^{-1} \frac{\partial}{\partial r} f \left( p\left( r, x \right) , T  \left( r \right) \right) = \mathcal{C} \left( p\left( r, x \right) , T  \left( r  \right) \right)  \, .
\end{align}
Since in this work DM production is mainly governed by $2 \leftrightarrow 2$ scattering processes we will discuss the collision term for these type of processes in more detail. For a $A + B \rightarrow C + DM$ scattering the collision term for the evolution of the momentum distribution function of DM is given by: 
\begin{align}
\mathcal{C_{DM}} \left( p \right) &= \frac{g_A g_B g_C}{2 E_{DM}} \int \frac{d^3 \mathbf{p}_A}{ 2 E_A \left( 2 \pi \right)^3} \frac{d^3 \mathbf{p}_B}{ 2 E_B \left( 2 \pi \right)^3} \frac{d^3 \mathbf{p}_C}{ 2 E_C \left( 2 \pi \right)^3} \left( 2 \pi \right)^4 \delta^{4}  \left( p_A + p_B - p_C - p_{DM} \right) \times \nonumber \\
 &\times |\mathcal{M}|^2 \left( f_A f_B - f_C f_{DM} \right) \, .
\end{align}
Here, $E_i = \sqrt{\mathbf{p}_i^2 + m_i^2}$, $\mathcal{M}$ is the matrix element for the process $A + B \rightarrow C + DM$ which is the same in both directions since we are assuming CP invariant interactions and $f_i$ is the distribution function of particle species $i$. We assume that $f_C f_{DM} \ll f_A f_B$ which is justified since the paper explores the freeze in production of DM. Furthermore, we take $f_{A/B}=f_{A/B}^{th}$ assuming the interactions of $A$ and $B$ are efficient enough to keep them in thermal equilibrium. Moreover, taking $f_{A/B}^{th}$ to be a Boltzmann distribution, shifting the integration over $\mathbf{p}_c$ to $\mathbf{p}_C + \mathbf{p}_{DM} = \mathbf{P}$ and multiplying the equation by $1 = \int dP_0 \delta \left( P_0 - E_C - E_{DM} \right)$ yields
\begin{align}
\mathcal{C} \left( p_{DM} \right) &= \frac{g_A g_B g_C}{4 E_{DM}} \int \frac{d^4 P}{\left( 2 \pi \right)^3} \frac{exp \left( -P_0 / T \right)}{E_C} \delta \left( P_0 - E_C - E_{DM} \right)  \times \nonumber \\
& \times \int \frac{d^3 \mathbf{p}_A}{ 2 E_A \left( 2 \pi \right)^3} \frac{d^3 \mathbf{p}_B}{ 2 E_B \left( 2 \pi \right)^3} \left( 2 \pi \right)^4 \delta^{4}  \left( p_A + p_B - p_C - p_{DM} \right) |\mathcal{M}|^2 \label{eq:interstep}
\end{align} 
The equation above can be simplified by rewriting it in terms of the reduced cross section \cite{Luty:1992un}: 
\begin{align}
&g_A g_B g_C g_{DM} \int \frac{d^3 \mathbf{p}_A}{ 2 E_A \left( 2 \pi \right)^3} \frac{d^3 \mathbf{p}_B}{ 2 E_B \left( 2 \pi \right)^3} \left( 2 \pi \right)^4 \delta^{4}  \left( p_A + p_B - p_C - p_{DM} \right) |\mathcal{M}|^2 \nonumber \\ = &\frac{\hat{\sigma} \left( s \right) }{\sqrt{\left[ 1 - \frac{\left( m_C + m_{DM} \right)^2}{s}  \right] \left[ 1 - \frac{\left( m_C - m_{DM} \right)^2}{s}  \right]}} \, .
\end{align}
Moreover, we change the variables of integration from $d^4 P$ to an integration over the zero component of the center of mass momentum vector $P_0$, the center of mass energy $s$ and the angle $\theta$ between center of mass momentum $\mathbf{P}$ and the momentum of the DM candidate $\mathbf{p}_{DM}$, $d^4 P = 2 \pi \mathbf{P^2} dP_0 d\mathbf{P} d\cos \left( \theta \right) = 2 \pi \sqrt{P_0^2 - s} dP_0 ds d\cos \left( \theta \right)$. To eliminate the remaining $\delta$ function we express the argument in terms of $\cos \left( \theta \right)$:
\begin{align}
\delta \left( E_C + E_{DM} - P_0 \right) &= \delta \left( \sqrt{\mathbf{P}^2 + \mathbf{p}_{DM}^2- 2 \mathbf{P} \mathbf{p}_{DM} \cos \left( \theta  \right) + m_C^2}   + E_{DM} - P_0 \right) \nonumber \\
&= \frac{E_C}{\mathbf{P} \mathbf{p}_{DM}} \delta \left(  \cos \left( \theta  \right) -  \cos \left( \theta_0  \right)\right) \, ,
\end{align}
where $\cos \left( \theta_0 \right)$ is the value required for $\cos \left( \theta \right)$ for a vanishing argument of the $\delta$ function. 
Therewith, eq. \eqref{eq:interstep} results in 
\begin{align}
\mathcal{C} \left( p_{DM} \right) &= \frac{1}{4 g_{DM} E_{DM} \mathbf{p}_{DM}} \int \limits_{s_\text{min}}^\infty ds \frac{\hat{\sigma} \left( s \right) }{\sqrt{\left[ 1 - \frac{\left( m_C + m_{DM} \right)^2}{s}  \right] \left[ 1 - \frac{\left( m_C - m_{DM} \right)^2}{s}  \right]}} \times \nonumber \\ &\times \int \limits_{\sqrt{s_\text{min}}}^\infty \frac{dP_0}{\left( 2 \pi  \right)^2} \exp \left( - \frac{P_0}{T} \right) \underbrace{\int \limits_{-1}^1 d\cos \left( \theta \right) \delta \left(  \cos \left( \theta  \right) -  \cos \left( \theta_0  \right)\right)}_\text{$=1$, if $\cos \left( \theta_0 \right) \in [-1,1]$} 
\end{align}
The last integral basically restricts the boundaries of either $P_0$ or $s$ in the sense that if
\begin{align}
\sqrt{\mathbf{P}^2 + \mathbf{p}_{DM}^2 - 2 \mathbf{P} \mathbf{p}_{DM} \cos \left( \theta_0  \right) + m_C^2}   + E_{DM} - P_0 = 0
\end{align}
is fulfilled $ |\cos \left( \theta_0  \right)| \leq 1$ must hold. This requirement yields the inequality
\begin{align}
\left( s + m_{DM}^2 - m_C^2 - 2 P_0 E_{DM} \right)^2 \leq 4 \mathbf{p}_{DM}^2 \left( P_0^2 - s \right) \, .
\end{align}
In case of $m_C = m_{DM}$ \footnote{This is a good approximation for this work since we assume the dark sector to be almost degenerate in mass.} this results in a lower (relative minus sign) and upper bound (relative plus sign) of the $P_0$ integration of
\begin{align}
P_0^{\pm} = \frac{E_{DM} s}{2 m_{DM}^2} \left[ 1 \pm \frac{p_{DM}}{E_{DM}} \sqrt{1 - 4 \frac{ m_{DM}^2}{ s}} \right] \overset{m_{DM} =0}{=} \left\{\begin{array}{ll} P_0^+ \rightarrow \infty\\
P_0^- = \frac{s}{4 p_{DM}} + p_{DM} \end{array}\right. \, . 
\end{align} 
The last equality is given to showcase that in case of $m_{DM} = 0$ only a lower bound exists, as was shown in \cite{Heeck:2017xbu}, while for finite DM masses there is also an upper bound. Thus, we have
\begin{align}
\mathcal{C} \left( p_{DM} \right) &= \frac{1}{4 g_{DM} E_{DM} \mathbf{p}_{DM}} \int \limits_{s_\text{min}}^\infty ds \frac{\hat{\sigma} \left( s \right) }{\sqrt{ 1 -4 \frac{ m_{DM}^2}{s}}} \int \limits_{P_0^-}^{P_0^+} \frac{dP_0}{\left( 2 \pi  \right)^2} \exp \left( - \frac{P_0}{T} \right) \, .
\end{align}
The $s$ integral and the following integration of the differential equation for an arbitrary cross section cannot be performed analytically. However, in case of a very light DM candidate ($m_{DM} \approx 0$) and a resonant production process with $\Gamma_\text{mediator} \ll M_\text{mediator}$ the integral can be evaluated analytically. Moreover, this case is of special interest for this work since for resonant production the DM mass turns out to be below $keV$. Therefore, the exact shape of the momentum distribution is required to quantify the impact of DM on structure formation. In this case we have $P_0^+ \rightarrow \infty$ and 
\begin{align}
\hat{\sigma} \left( s \right) \approx \delta \left( s - M_N^2 \right) \sqrt{1- 4 \frac{m_{DM}^2}{s}} \hat{\sigma}_{BW} \left( s \right) \, .
\end{align}
Hence the collision term yields 
\begin{align}
\mathcal{C} \left( p_{DM} \right) &= \frac{T}{32 \pi^2 g_{DM} \mathbf{p}_{DM}^2} \hat{\sigma}_{BW} \left( M_N^2 \right) \exp \left( - \frac{M_N^2}{4 \mathbf{p}_{DM} T} - \frac{\mathbf{p}_{DM}}{T} \right) \, .
\end{align}
Transforming the variables according to eq. \eqref{eq:r} and eq. \eqref{eq:x} and taking $g_s$ to be a constant, i.e. $x = \frac{\mathbf{p}_{DM}}{T}$, leads to
\begin{align}
\mathcal{C} \left(p_{DM} \right) = \frac{1}{32 \pi^2 g_{DM}} \frac{r}{x^2 m_0} \hat{\sigma}_{BW} \left( M_N^2 \right) \exp \left( - \frac{M_N^2 r^2}{4 x m_0^2} - x \right) \, .
\end{align}
A collision term of this form can be integrated and results in the following momentum distribution function:
\begin{align}
f \left( p , T  \right) = \frac{M_{pl} \hat{\sigma}_{BW} \left( M_N^2 \right)}{64 \pi^2 g_{DM} c_H} \frac{\exp \left( - p / T \right)}{M_N^3} \frac{T^2}{p^2} \left[ \sqrt{\frac{\pi p}{T}} \text{erf} \left( \frac{M_N}{\sqrt{p T}} \right) - 2 \frac{M_N}{T} \exp \left( - \frac{M_N^2}{T p} \right) \right] \, , \label{eq:mdf}
\end{align}
where $\text{erf} \left( x \right)$ is the error function. Therewith, the number density is given by the integration over the momentum
\begin{align}
n \left( T \right) = 4 \pi g_{DM} \int \limits_0^\infty p^2 f \left( p , T \right)  \overset{T \ll M_N}{=} \frac{M_{pl} \hat{\sigma}_{BW} \left( M_N^2 \right)}{8 c_H} \frac{T^3}{M_N^3} \, . \label{eq:resultapp}
\end{align}
In the last step, we assumed that the temperature where we observe the DM density is much smaller than the mass of the resonant particle. As mentioned above, to derive this analytic result we took the effective entropy degrees of freedom to be a constant. Hence the above formula is only a good approximation as long as we take $T$ large enough to stay at a constant value of $g_s \left( T \right) \approx 100$. Of course, we observe the universe at a smaller temperature. However, the above result remains a good approximation if the main part of the production has been finished before $g_s \left( T \right)$ starts to vary significantly since for a collisionless particle species the quantity $Y=\frac{n}{s}$ is a constant. \\
By comparing the number of produced DM particles at temperature $T$ to the number of particles for $T \rightarrow 0$, $\frac{n \left( T \right) T^3}{\underset{T \rightarrow 0}{\text{lim } }n \left( T \right) T^3}$, with an unapproximated $n \left( T \right)$ we find that for $T \approx \frac{M_N}{4}$ already over $0.99$ of DM particle have been produced. Thus, as long as $M_N \geq 100 \, \mathrm{GeV}$ the result \eqref{eq:resultapp} serves as a good estimate. \\
Beside collision terms for $2 \leftrightarrow 2$ scattering processes, the collision term for the (inverse) decay $N \leftrightarrow \nu h$ is required. The procedure for performing the integration over the particle momenta follows the same lines as for the $2 \leftrightarrow 2$ scattering. Thus, we only give the result for the collision term resulting from the decay that appears in the Boltzmann equation for the heavy neutrino $N$:
\begin{align}
\mathcal{C}_N \left( p_N \right) = \frac{M_N}{\sqrt{p_N^2 + M_N^2}} \left[ \frac{y_\nu^2 g_\nu g_h}{16 \pi} M_N \exp \left( - \frac{\sqrt{p_N^2 + M_N^2}}{T} \right) - \Gamma_{N \rightarrow \nu h} f_N \left( p_N , T \right) \right] \, . \label{eq:ColN}
\end{align} 
%%%%%%%%%%%%%%%%%%%%%%%%%%%%%%%%%%%%%%%
\section{Cross Sections} \label{app:xsection}
Here, we give the relevant reduced cross sections for the case $m_\phi = m_\chi$. Since CP conservation is assumed the reduced cross sections for a process and its time reserved process are the same. 
\begin{align}
\hat{\sigma}_{v_i h \leftrightarrow \chi \phi} \left( s \right) &= \left( \sum_j \left( Y_\nu \right)_{ij} y_\chi \right)^2 \frac{\left( 1 - \frac{m_h^2}{s} \right)^2}{32 \pi} \frac{s^2 \sqrt{1 - 4 \frac{m_\chi^2}{s}}}{\left(s - M_N^2 \right)^2 + \Gamma_N^2 M_N^2} \label{eq:SMScattering}
\end{align}
Here, $\Gamma_N$ is the total decay width of the propagating neutrino which can decay into $v h$ for $M_N > m_h$ and into $\chi \phi$ for $M_N > 2 m_\chi$. The decay width is given by:
\begin{align}
\Gamma_N &=  y_\nu^2 \frac{\left( M_N^2 - m_h^2 \right)^2}{16 \pi M_N^3} + y_\chi^2 \frac{\left( M_N + 2 m_\chi \right) \sqrt{M_N^2 - 4 m_\chi^2}}{16 \pi M_N} \, .  \label{eq:Decay}
\end{align}
\begin{align}
 \sigma_{Wl \rightarrow \chi \phi} &= y_\chi^2 y_\nu^2 \frac{3 M_W^2}{24 \pi s M_N \left( s - M_N^2 \right)^2} \left[ \left( M_W^2 - m_l^2 \right) \left( M_W^2 + 2 \left( m_l^2 - M_W^2 \right) - 4 M_N m_\chi \right) \right. \nonumber \\
 &\left. + \left( M_N^2+ m_l^2 - M_W^2 + 4M_N m_\chi \right) \right] \sqrt{\frac{s \left( s - 4 m_\chi^2 \right)}{m_l^4 + \left( s - M_W^2 \right)^2 - 2m_l^2 \left( s + M_W^2 \right)}}
\end{align}
\begin{align}
\sigma_{Z \nu \rightarrow \chi \phi} &=y_\chi^2 y_\nu^2 \frac{3 M_W^2 \sqrt{1 - \frac{4 m_\chi^2}{s}}}{16 \pi^2 M_N^2 \left( s- M_N^2 \right) \left( s- M_Z^2 \right)} \left[ \left( s + M_Z^2 \right) M_N^2 + 4 M_N m_\chi \left( s - M_Z^2 \right) \nonumber \right. \\
 &\left. + s^2 - s M_Z^2 - 2 M_Z^4 \right] 
\end{align}
\begin{align}
\hat{\sigma}_{NN \rightarrow \chi \chi} &= \frac{y_\chi^4}{32 \pi s } \left[ \frac{\sqrt{\left( s - 4m_\chi^2 \right) \left( s - 4 M_N^2 \right)} \left(2 M_N^4 - 4 M_N^2 m_\chi^2 + m_\chi^2 s \right)}{M_N^4 - 4 M_N^2 m_\chi^2 + m_\chi^2 s} \right. \nonumber \\  
&\left.- 4M_N^2 \arccoth \left( \frac{2 M_N^2 -s}{\sqrt{\left( s - 4m_\chi^2 \right) \left( s - 4 M_N^2 \right)}} \right) \right]  \label{eq:HN1}
\end{align}
\begin{align}
 \sigma_{N N \rightarrow \phi \phi } &= y_\chi^4 \left( 1 - \frac{4 m_\chi^2}{s} \right) \left[ - \sqrt{\left( s - 4 M_N^2 \right) \left( s - 4m_\chi^2 \right)} \left( m_\chi^2 s + 2 M_N^4 + 4 M_N^3 m_\chi \right)  \right. \nonumber \\
 &\left.+  2 \left[ 2 M_N \left( 2 m_\chi M_N \right) + s \right] \left[ m_\chi^2 \left( s - 4 M_N^2 \right) + M_N^4 \right] \nonumber \right. \\
 &\left. \times \arctanh \left( \frac{\sqrt{\left( s - 4 M_N^2 \right) \left( s - 4m_\chi^2 \right)}}{s - 2 M_N^2} \right)  \right]
\end{align}

\bibliography{references}
\bibliographystyle{h-physrev}

\end{document}